\documentclass[prb,twocolumn,amsmath,amssymb,apsrev]{revtex4}

\usepackage{graphicx} % Include figure files
\usepackage{dcolumn} % Align table columns on decimal point
\usepackage{color} % \textcolor for comments
\usepackage{dsfont}
\def\ba{{\bf a}}
\def\bA{{\bf A}}
\def\bB{{\bf B}}
\def\bE{{\bf E}}

\def\bk{{\bf k}}

\def\bp{{\bf p}}
\def\bq{{\bf q}}
\def\br{{\bf r}}

\def\bQ{{\bf Q}}

\def\b0{{\bf 0}}

\def\cA{{\cal A}}
\def\cC{{\cal C}}
\def\cE{{\cal E}}
\def\cF{{\cal F}}
\def\cG{{\cal G}}
\def\cH{{\cal H}}
\def\cM{{\cal M}}
\def\cN{{\cal N}}
\def\cP{{\cal P}}

\def\Re{{\rm Re}}
\def\Im{{\rm Im}}
\def\bra{\langle}
\def\ket{\rangle}
\def\up{\uparrow}
\def\down{\downarrow}
\def\lra{\leftrightarrow}

\def\alf{\alpha}
\def\eps{\epsilon}
\def\gam{\gamma}
\def\Gam{\Gamma}
\def\lam{\lambda}

\def\om{\omega}
\def\Om{\Omega}
\def\sg{\sigma}

\def\sgn{{\rm sgn}}

\def\tlam{\tilde\lambda}

\def\trp{{\rm Tr}_p}
\def\widetrp{\widetilde{\rm Tr}_p}

\def\Kabgd{K^{\alf\beta\gam\delta}_{iq_0}}

\def\sgc{{\cal S_\bp}}

\begin{document}

\title{Longitudinal conductivity and Hall coefficient \\ in two-dimensional metals with spiral magnetic order}
%\title{Electromagnetic response of two-dimensional metals with spiral magnetic order}

\author{Johannes Mitscherling}
\affiliation{Max Planck Institute for Solid State Research,
 D-70569 Stuttgart, Germany}
\author{Walter Metzner}
\affiliation{Max Planck Institute for Solid State Research,
 D-70569 Stuttgart, Germany}

\date{\today}

\begin{abstract}
 We compute the longitudinal dc conductivity and the Hall conductivity in a two-dimensional metal with spiral magnetic order. Scattering processes are modeled by a momentum-independent relaxation rate $\Gam$. We derive expressions for the conductivities, which are valid for arbitrary values of $\Gam$. Both intraband and interband contributions are fully taken into account. For small $\Gam$, the ratio of interband and intraband contributions is of order $\Gam^2$. In the limit $\Gam \to 0$, the conductivity formulas assume a simple quasiparticle form, as derived by Voruganti {\it et al.} [Phys. Rev. B {\bf 45}, 13945 (1992)]. Using the complete expressions, we can show that relaxation rates in the regime of recent transport experiments for cuprate superconductors in high magnetic fields are sufficiently small to justify the application of these simplified formulas. The longitudinal conductivity exhibits a pronounced nematicity in the spiral state. The drop of the Hall number as a function of doping observed recently in several cuprate compounds can be described with a suitable phenomenological ansatz for the magnetic order parameter.
\end{abstract}

\maketitle

\newpage

%%% Intro %%%%%%%%%%%%%%%%%%%%%%%%%%%%%%%%%%%%%%%%%%%%%%%%%%%%%%%%%%%%%%%%

\section{Introduction}

Understanding the ``normal'' ground state in the absence of superconductivity is the key to 
understanding the fluctuations that govern the anomalous behavior of cuprate superconductors above the critical temperature. \cite{broun08} Superconductivity can be suppressed by applying a magnetic field, but very high fields are required for a complete elimination in high-temperature superconductors. Recently, magnetic fields up to 88 T were achieved, such that the critical temperature of $\rm YBa_2Cu_3O_y$ (YBCO) and other cuprate compounds could be substantially suppressed even at optimal doping.
Charge transport measurements in such high magnetic fields indicate a drastic reduction of the charge-carrier density in a narrow doping range upon entering the pseudogap regime. \cite{badoux16,collignon17} In particular, Hall measurements at various dopings $p$ yield a drop of the Hall number from $1+p$ to values near $p$. 

The drop in carrier density indicates a phase transition associated with a Fermi-surface reconstruction. Storey \cite{storey16} has pointed out that the observed Hall number drop is consistent with the formation of a N\'eel antiferromagnet, and also with a Yang-Rice-Zhang (YRZ) state \cite{yang06} without long-range order. Other possibilities are fluctuating antiferromagnets, \cite{qi10} fractionalized Fermi liquids, \cite{chatterjee16} and charge density wave states. \cite{caprara17}
As long as no spectroscopic measurements are possible in high magnetic fields, it is hard to confirm or rule out any of these candidates experimentally.
For strongly underdoped cuprates, where superconductivity is absent or very weak, neutron scattering probes show that the N\'eel state is quickly destroyed upon doping, in agreement with theoretical findings. For underdoped YBCO incommensurate antiferromagnetic order has been observed. \cite{haug09,haug10}

On the theoretical side, microscopic calculations yield incommensurate antiferromagnetism as the most robust order parameter in the non-superconducting ground state over a wide doping range away from half filling. For the two-dimensional Hubbard model, antiferromagnetic order with wave vectors $\bQ$ away from the N\'eel point $(\pi,\pi)$ was found in numerous mean-field calculations, \cite{schulz90,kato90,fresard91,raczkowski06,igoshev10} and also by expansions for small hole density, where fluctuations are taken into account.\cite{chubukov92,chubukov95}
At weak coupling magnetic order with $\bQ \neq (\pi,\pi)$ was confirmed by functional renormalization group calculations, \cite{metzner12,yamase16} and at strong coupling by state-of-the-art numerical techniques. \cite{zheng17} Recent dynamical mean-field calculations with vertex corrections suggest that the Fermi-surface geometry determines the (generally incommensurate) ordering wave vector not only at weak coupling, but also at strong coupling. \cite{vilardi18} 
For the two-dimensional $t$-$J$ model, the strong-coupling limit of the Hubbard model, expansions for small hole density indicate that the N\'eel state is stable only at half filling, and is replaced by a spiral antiferromagnet upon doping. \cite{shraiman89,kotov04}

There is a whole zoo of distinct magnetic states. The most favorable, or at least the most popular, are collinear states, combined with charge order to form spin-charge stripes, and planar spiral states. Stripe order has been observed in $\rm La$-based cuprates. \cite{tranquada95} Theoretically, commensurate stripe order was shown to minimize the ground-state energy of the strongly interacting Hubbard model with pure nearest-neighbor hopping at doping $1/8$. \cite{zheng17} However, this is a very special choice of parameters, and stripe order is not ubiquitous in cuprates.
Recently, it was shown that it is difficult to explain the recent high-field transport experiments in cuprates by collinear magnetic order. \cite{charlebois17}
Generally, the energy difference between different magnetic states seems to be rather small.

In the present paper we compute the electrical conductivity and the Hall coefficient for planar spiral magnetic states. In a spiral magnet, the electron band is split in only two quasiparticle bands. In this respect, the spiral state is as simple as the N\'eel state. By contrast, all other magnetically ordered states entail a fractionalization in many subbands, actually infinitely many in the case of incommensurate order. Hence, only the spiral magnet forms a metal with a simple Fermi surface, defined by the border of a small number of electron and hole pockets.
For a sufficiently large order parameter there are only hole pockets in the hole-doped system. The spectral weight for single-electron excitations is strongly anisotropic, so that the spectral function exhibits Fermi arcs, \cite{eberlein16} which are a characteristic feature of the pseudogap phase in high-$T_c$ cuprates.

The electromagnetic response of spiral magnetic states has already been analyzed by Voruganti {\it et al.} \cite{voruganti92} They derived formulas for the dc conductivity and for the Hall conductivity in the low-field limit, that is, to linear order in the magnetic field. The spiral states were treated in mean-field approximation. The resulting expressions have the same form as for non interacting electrons, with the bare dispersion relation replaced by quasiparticle bands. Assuming a simple phenomenological form of the spiral order parameter as a function of doping, Eberlein {\it et al.}\ \cite{eberlein16} showed that the Hall conductivity computed with the formula derived by Voruganti {\it et al.}\ indeed exhibits a drop of the Hall number consistent with the recent experiments \cite{badoux16,collignon17} on cuprates in high magnetic fields. Most recently an expression for the thermal conductivity in a spiral state has been derived along the same lines, and a similar drop in the carrier density has been found. \cite{chatterjee17}

The expressions for the electrical and Hall conductivities derived by Voruganti {\it et al.}\ \cite{voruganti92} have been obtained for small relaxation rates. However, the relaxation rate in the cuprate samples studied experimentally is sizable. For example, in spite of the high magnetic fields, the product of cyclotron frequency and relaxation time $\om_c \tau$ extracted from the experiments on $\rm La_{1.6-x} Nd_{0.4} Sr_x Cu O_4$ (Nd-LSCO) samples is as low as $0.075$. \cite{collignon17}
Moreover, from the derivation of Voruganti {\it et al.}, the precise criterion for a ``small'' relaxation rate is not clear. Hence, we derive complete expressions for the electrical and Hall conductivities allowing for relaxation rates of arbitrary size. We assume $\om_c \tau \ll 1$ such that an expansion to linear order in the magnetic field is indeed sufficient. We find that the relatively simple formulas derived by Voruganti {\it et al.}\ are valid only if the relaxation rate is much smaller than the direct gap between the upper and the lower quasiparticle band. Otherwise interband terms yield additional contributions. For a sizable relaxation rate the drop in the Hall number caused by the magnetic order is less steep than for a small relaxation rate. However, a numerical evaluation of the conductivity formulas shows that for parameters relevant for cuprate superconductors, the interband contributions play only a minor role.
Applying the conductivity formulas to cuprates we show that the longitudinal conductivities exhibit a pronounced nematicity in the spiral state, and the observed Hall number drop can be fitted with realistic parameters.

This paper is structured as follows. In Sec.~II we derive the formulas for the dc conductivity and the Hall conductivity. We recapitulate the formalism provided already by Voruganti {\it et al.}, \cite{voruganti92} for the sake of a coherent notation and a self-contained presentation. Lengthy algebra is carried out in the appendixes. In Sec.~III we discuss results with a focus on the role of interband terms in the presence of a sizable relaxation rate. Here we make contact to the recent experiments in cuprates. A conclusion in Sec.~IV closes the presentation.

%%% Formalism %%%%%%%%%%%%%%%%%%%%%%%%%%%%%%%%%%%%%%%%%%%%%%%%%%%%%%%%%%%%

\section{Formalism}

%%%%%%%%%%%%%%%%%%%%%%%%%%%%%%%%%%%%%%%%%%%%%%%%%%%%%%%%%%%%%%%%%%%%%%%%%%

\subsection{Action}

We begin by recapitulating the derivation of the action describing electrons with spiral magnetic order coupled to an electromagnetic field.\cite{voruganti92}
We use natural units such that $\hbar = 1$ and $c=1$.
The kinetic energy of electrons in a tight-binding representation of a single valence band is given by
\begin{equation} \label{H_0}
 H_0 = \sum_{j,j'} \sum_{\sg} t_{jj'} c_{j,\sg}^\dag c_{j',\sg}^{\phantom\dag} =
 \sum_{\bp,\sg} \eps_\bp \, a_{\bp,\sg}^\dag a_{\bp,\sg}^{\phantom\dag} \, ,
\end{equation}
where $t_{jj'}$ is the hopping amplitude between lattice sites $j$ and $j'$. The momentum-dependent band energy $\eps_\bp$ is the Fourier transform of $t_{jj'}$. The operators $c$ and $c^\dag$ are electron annihilation and creation operators in real space, respectively, while $a$ and $a^\dag$ are the corresponding operators in momentum space. The index $\sg$ describes the spin orientation $\up$ or $\down$.

The mean-field Hamiltonian for electrons in a spiral magnetic state has the form \cite{fresard91,igoshev10}
\begin{equation} \label{H}
 H = H_0 - \sum_\bp \Delta \left(
 a_{\bp+\bQ/2,\up}^\dag a_{\bp-\bQ/2,\down}^{\phantom\dag}
 + a_{\bp-\bQ/2,\down}^\dag a_{\bp+\bQ/2,\up}^{\phantom\dag} \right) \, .
\end{equation}
The ``magnetic gap'' $\Delta$ is a convenient order parameter quantifying the strength of the spiral order. It can be chosen real. In a mean-field solution of the Hubbard model, the magnetic gap is defined by
$\Delta = U \bra a_{\bp+\bQ/2,\up}^\dag a_{\bp-\bQ/2,\down}^{\phantom\dag} \ket$.
We have dropped a constant in Eq.~(\ref{H}) which contributes to the total energy, but not to the electromagnetic response.

The action corresponding to the Hamiltonian $H$ can be written most conveniently by using Grassmann spinors of the form
\begin{equation}
 \Psi_p =
 \left( \begin{array}{l} \psi_{ip_0,\bp+\bQ/2,\up} \\
 \psi_{ip_0,\bp-\bQ/2,\down} \end{array} \right) \, .
\end{equation}
Here and in the following we use frequency-momentum variables $p = (ip_0,\bp)$, where $p_0$ is a 
fermionic Matsubara frequency. 
The action can then be written as
\begin{equation}
 {\cal S}[\Psi,\Psi^*] = - \sum_p \Psi_p^* G_p^{-1} \Psi_p \, ,
\end{equation}
with the inverse matrix propagator given by
\begin{equation}
 G_p^{-1} = \left( \begin{array}{cc}
 ip_0 + \mu - \eps_{\bp+\bQ/2} & \Delta \\
 \Delta & ip_0 + \mu - \eps_{\bp-\bQ/2} \end{array} \right) \, ,
\end{equation}
where $\mu$ is the chemical potential.

We now couple the system to electromagnetic fields. We choose a gauge such that the scalar potential vanishes. The electric and magnetic fields are thus entirely determined by the vector potential $\bA(\br,t)$ as
$\bE(\br,t) = - \partial_t\bA(\br,t)$ and $\bB(\br,t) = \nabla \!\times\! \bA(\br,t)$, respectively.
The coupling to the orbital motion of the electrons gives rise to a phase factor multiplying the hopping amplitudes \cite{wilson74}
\begin{equation} \label{t_A}
 t_{jj'}[\bA] =
 t_{jj'} \exp \Big( ie \int_{\br_{j'}}^{\br_j} \bA(\br,t) \cdot d\br \Big) \, ,
\end{equation}
where $e<0$ is the electron charge, and $\br_j$ is the spatial position of the lattice site labeled by $j$. We discard the Zeeman coupling of the magnetic field to the electron spin, since it has no significant effect on our results. 
For fields varying slowly between lattice sites connected by $t_{jj'}$, which is the case we are interested in, we can parametrize $\bA(\br,t)$ by a link variable $\bA_{jj'}(t) = \bA[(\br_j + \br_{j'})/2,t]$, and approximate the line integral in Eq.~(\ref{t_A}) by
\begin{equation}
 \int_{\br_{j'}}^{\br_j} \bA(\br,t) \cdot d\br = \bA_{jj'}(t) \cdot \br_{jj'} \, ,
\end{equation}
with $\br_{jj'} = \br_j - \br_{j'}$.
The action is expressed in terms of the Wick-rotated vector potential, that is, for imaginary times $\tau$, which we denote by $\bA_{jj'}(\tau)$.
Expanding Eq.~(\ref{t_A}) in powers of $\bA$, one obtains the complete action \cite{voruganti92}
\begin{equation} \label{S_A}
 {\cal S}[\Psi,\Psi^*;\bA] = - \sum_p \Psi_p^* G_p^{-1} \Psi_p +
 \sum_{p,p'} \Psi_p^* V_{pp'}[\bA] \Psi_{p'} \, ,
\end{equation}
where the coupling to the electromagnetic field has the form
\begin{equation} \label{V_A}
 V_{pp'}[\bA] = \sum_{n=1}^\infty \frac{e^n}{n!} \sum_{q_1,\dots,q_n}
 \lam_{\bp\bp'}^{\alf_1 \dots \alf_n} \, 
 A_{q_1}^{\alf_1} \dots A_{q_n}^{\alf_n} \,
 \delta_{p-p',\sum_{i=1}^n q_i} \, .
\end{equation}
Here and in the following we use Einstein's summation convention for repeated Greek indices.
$A_q^\alf$ with $\alf=x,y,z$ and $q = (iq_0,\bq)$ is the $\alf$ component of the Fourier transform $\bA_q$ of $\bA_{jj'}(\tau)$, that is,
\begin{equation}
 \bA_{jj'}(\tau) = \sum_{q} \bA_q \,
 e^{i\left[\frac{1}{2} \bq (\br_j + \br_{j'}) - q_0\tau \right]} \, .
\end{equation}
The $n$th-order vertices are given by
\begin{equation} \label{lampp'}
 \lam_{\bp\bp'}^{\alf_1 \dots \alf_n} = \left( \begin{array}{cc}
 \eps_{\bp/2+\bp'/2+\bQ/2}^{\alf_1 \dots \alf_n} & 0 \\
 0 & \eps_{\bp/2+\bp'/2-\bQ/2}^{\alf_1 \dots \alf_n} \end{array} \right) \, ,
\end{equation}
where
$\eps_\bp^{\alf_1 \dots \alf_n} =
 \partial^n \eps_\bp/(\partial p_{\alf_1} \dots \partial p_{\alf_n})$.
Current-relaxing scattering processes are taken into account in the simplest fashion by adding a fixed relaxation rate $\Gam$ to the (inverse) propagator, such that
\begin{widetext}
\vspace{-0.6cm}
\begin{equation} \label{G_p}
 G_p^{-1} = \left( \begin{array}{cc}
 ip_0 + \mu - \eps_{\bp+\bQ/2} + i\Gam\sgn(p_0) & \Delta \\[1mm] 
 \Delta & ip_0 + \mu - \eps_{\bp-\bQ/2} + i\Gam\sgn(p_0) \end{array} \right) \, .
\end{equation}
\vspace{-0.1cm}
\end{widetext}
A relaxation term of this form is obtained, for example, from scattering at short-ranged impurity potentials in Born approximation.\cite{rickayzen80}

%%%%%%%%%%%%%%%%%%%%%%%%%%%%%%%%%%%%%%%%%%%%%%%%%%%%%%%%%%%%%%%%%%%%%%%%%%%%%%

\subsection{Current and response functions}

The action ${\cal S}[\Psi,\Psi^*;\bA]$ in Eq.~(\ref{S_A}) is quadratic in the fermion fields. The partition function is thus given by a Gaussian integral. Performing the integral, taking the logarithm, and expanding in powers of $V[\bA]$ yields the grand canonical potential in the form \cite{voruganti92}
\begin{equation} \label{OmA}
 \Om[\bA] = \Om_0 + T \sum_{n=1}^\infty \frac{1}{n} {\rm Tr} (GV[\bA])^n \, ,
\end{equation}

where $T$ is the temperature, and $\Om_0$ is the grand canonical potential in the absence of $\bA$. Both $G$ and $V[\bA]$ are matrices in frequency-momentum and spin space, where $G_{pp'} = \delta_{pp'} G_p$ with $G_p$ given by Eq.~(\ref{G_p}), and $V_{pp'}[\bA]$ is given by Eq.~(\ref{V_A}). The trace is a sum over frequency, momentum, and spin orientation. Note that $V_{pp'}[\bA]$ is diagonal in spin space, since the electromagnetic field couples only to the charge.

The charge current is defined by the first derivative of the grand canonical potential with respect to the vector potential,
\vspace{-0.12cm}
\begin{equation} \label{jq}
 j_q^\alf = - \frac{1}{L} \frac{\partial\Om[\bA]}{\partial A_{-q}^\alf} \, ,
\end{equation}
where $L$ is the number of lattice sites. We choose units of length such that a single lattice cell has volume 1, so that $L$ corresponds to the volume of the system.
An expansion of the current in powers of $\bA$ has the general form
\begin{equation} \label{jqexp}
 j_q^\alf = - \sum_{n=1}^\infty \sum_{q_1,\dots,q_n}
 K_{q_1 \dots q_n}^{\alf\alf_1\dots\alf_n} A_{q_1}^{\alf_1} \dots A_{q_n}^{\alf_n}
 \delta_{q,q_1+\dots+q_n} \, .
\end{equation}

In this work we focus on the dc conductivity $\sg^{\alf\beta}$ and the dc Hall conductivity $\sg_H^{\alf\beta\gam}$, which describe the current response to homogeneous static electric and magnetic fields,
\begin{equation}
 j^\alf =
 \left[ \sg^{\alf\beta} + \sg_H^{\alf\beta\gam} B^\gam \right] E^\beta \, ,
\end{equation}
to leading order in $\bE$ and $\bB$.
In a microscopic calculation the dc response is obtained as the zero-frequency limit of the response to a spatially homogeneous dynamical electric field $\bE(t)$. A constant magnetic field is associated with a vector potential depending only on space, not on time. Following Voruganti {\it et al.},\cite{voruganti92} we therefore split the vector potential as $\bA(\br,t) = \ba^E(t) + \ba^B(\br)$, such that $\bE(t) = -\partial_t \, \ba^E(t)$ and $\bB(\br) = \nabla \times \ba^B(\br)$.
The corresponding Fourier transforms are related by $\bE_\om = i\om \, \ba_\om^E$ and $\bB_\bq = i \bq \!\times\! \ba_\bq^B$.

For imaginary time fields of the form $\bA(\br,\tau) = \ba^E(\tau) + \ba^B(\br)$, the expansion (\ref{jqexp}), carried out to the relevant order, can be written as
\begin{equation} \label{jq_exp}
 j_q^\alf = - \left( \delta_{\bq,\b0} K_{iq_0}^{\alf\beta} +
 K_{\bq,iq_0}^{\alf\beta\gam} a_\bq^{B,\gam} \right) a_{iq_0}^{E,\beta} + \dots \, ,
\end{equation}
where $a_{iq_0}^{E,\beta}$ is the temporal Fourier transform of $a^{E,\beta}(\tau)$, and $a_\bq^{B,\gam}$ is the spatial Fourier transforms of $a^{B,\gam}(\br)$.
We denote the analytic continuation of $K_{iq_0}^{\alf\beta}$ and
$K_{\bq,iq_0}^{\alf\beta\gam}$ to real frequencies, $iq_0 \to \om + i0^+$, by
$K_{\om}^{\alf\beta}$ and $K_{\bq,\om}^{\alf\beta\gam}$, respectively.
The dc conductivity is obtained as the zero frequency limit of the dynamical conductivity $\sg_\om^{\alf\beta}$, which is related to $K_\om^{\alf\beta}$ by
\begin{equation} \label{Ksg}
 K_\om^{\alf\beta} = -i\om \sg_\om^{\alf\beta} \, .
\end{equation}
The Hall conductivity is obtained from the zero frequency and zero momentum limit of the dynamical quantity $\sg_{H,\bq,\om}^{\alf\beta\eta}$, related to
$K_{\bq,\om}^{\alf\beta\gam}$ by
\begin{equation} \label{Ksg_H}
 K_{\bq,\om}^{\alf\beta\gam} =
 - \om \eps^{\gam\delta\eta} q^\delta \sg_{H,\bq,\om}^{\alf\beta\eta} \, .
\end{equation}
%

%%%%%%%%%%%%%%%%%%%%%%%%%%%%%%%%%%%%%%%%%%%%%%%%%%%%%%%%%%%%%%%%%%%%%%%%%%%%%%%%%%%%%%%

\subsection{Diagonalization}

To evaluate the trace in Eq.~(\ref{OmA}), it is convenient to use a basis in which $G_p$ is diagonal. This can be achieved by the unitary transformation \cite{voruganti92}
\begin{equation}
 U_\bp = \left( \begin{array}{rr} \cos\theta_\bp & \sin\theta_\bp \\
                - \sin\theta_\bp & \cos\theta_\bp \end{array} \right) \, ,
\end{equation}
where the rotation angle $\theta_\bp$ must satisfy the condition
\begin{equation}
 \tan(2\theta_\bp) = \frac{2\Delta}{\eps_{\bp+\bQ/2} - \eps_{\bp-\bQ/2}} \, .
\end{equation}
The transformed propagator has the diagonal form
\begin{align} \label{cG-1}
 \cG_p^{-1} &= U_\bp^\dag G_p^{-1} U_\bp \nonumber \\[1mm] =
 &\left( \begin{array}{cc} ip_0 + i\Gam\sgn(p_0) - E_\bp^+ & 0 \\
        0 & ip_0 + i\Gam\sgn(p_0) - E_\bp^- \end{array} \right),
\end{align}
where $E_\bp^+$ and $E_\bp^-$ are the two quasiparticle bands in the spiral magnetic state. Their momentum dependence is given by
%\begin{widetext}
%The transformed propagator has the diagonal form
%%
%\begin{equation} \label{cG-1}
% \cG_p^{-1} = U_\bp^\dag G_p^{-1} U_\bp =
% \left( \begin{array}{cc} ip_0 + i\Gam\sgn(p_0) - E_\bp^+ & 0 \\
%        0 & ip_0 + i\Gam\sgn(p_0) - E_\bp^- \end{array} \right) \, ,
%\end{equation}
%%
%where $E_\bp^+$ and $E_\bp^-$ are the two quasiparticle bands in the spiral magnetic %state. Their momentum dependence is given by
%\end{widetext}
%
\begin{equation} \label{Ep}
 E_\bp^\pm = g_\bp \pm \sqrt{h_\bp^2 + \Delta^2} - \mu \, ,
\end{equation}
where $g_\bp = \frac{1}{2} \big(\eps_{\bp+\bQ/2} + \eps_{\bp-\bQ/2}\big)$ and $h_\bp = \frac{1}{2} \big(\eps_{\bp+\bQ/2} - \eps_{\bp-\bQ/2}\big)$.

The vertices $\lam_{\bp\bp'}^{\alf_1 \dots \alf_n}$ have to be transformed accordingly as
\begin{equation}
 \tilde\lam_{\bp\bp'}^{\alf_1 \dots \alf_n} =
 U_\bp^\dag \lam_{\bp\bp'}^{\alf_1 \dots \alf_n} U_{\bp'} \, . 
\end{equation}
In the following we will mainly deal with the first- and second-order vertices for vanishing  momentum transfer, that is, $\bp = \bp'$.
The rotated first-order vertex has the form \cite{voruganti92}
\begin{equation}
 \tilde\lam_{\bp\bp}^{\alf} = 
 \left( \begin{array}{ll} E_\bp^{+,\alf} & F_\bp^\alf \\
 F_\bp^\alf & E_\bp^{-,\alf} \end{array} \right) \, ,
\end{equation}
where
$E_\bp^{\pm,\alf} = \partial E_\bp^\pm/\partial p_\alf$, and
\begin{equation} \label{Fp}
 F_\bp^\alf = \frac{2\Delta}{E_\bp^+ - E_\bp^-} h_\bp^\alf \, ,
\end{equation}
with $h_\bp^\alf = \partial h_\bp/\partial p_\alf$.
The rotated second-order vertex can be written as \cite{voruganti92}
\begin{equation}
 \tilde\lam_{\bp\bp}^{\alf\beta} = 
 \left( \begin{array}{cc}
 E_\bp^{+,\alf\beta} - C_\bp^{\alf\beta} & H_\bp^{\alf\beta} \\
 H_\bp^{\alf\beta} & E_\bp^{-,\alf\beta} + C_\bp^{\alf\beta}
 \end{array} \right) \, ,
\end{equation}
where $E_\bp^{\pm,\alf\beta} = \partial^2 E_\bp^\pm/(\partial p_\alf \partial p_\beta)$,
\begin{equation} \label{Cp}
 C_\bp^{\alf\beta} = \frac{2}{E_\bp^+ - E_\bp^-} F_\bp^\alf F_\bp^\beta =
 \frac{8 \Delta^2}{(E_\bp^+ - E_\bp^-)^3} h_\bp^\alf h_\bp^\beta \, ,
\end{equation}
and
\begin{equation} \label{Hp}
 H_\bp^{\alf\beta} = \frac{2\Delta}{E_\bp^+ - E_\bp^-} h_\bp^{\alf\beta} \, ,
\end{equation}
with $h_\bp^{\alf\beta} = \partial^2 h_\bp/(\partial p_\alf \partial p_\beta)$.

In subsequent derivations it will be convenient to decompose the vertices in diagonal and off-diagonal parts, such as
$\tilde\lam_{\bp\bp}^{\alf} = \cE_\bp^\alf + \cF_\bp^\alf$, with
\begin{equation} \label{cEcF}
 \cE_\bp^\alf = \left( \begin{array}{cc} E_\bp^{+,\alf} & 0 \\
 0 & E_\bp^{-,\alf} \end{array} \right) , \;
 \cF_\bp^\alf = \left( \begin{array}{cc} 0 & F_\bp^\alf \\
 F_\bp^\alf & 0 \end{array} \right) ,
\end{equation}
and $\tilde\lam_{\bp\bp}^{\alf\beta} =
\cE_\bp^{\alf\beta} - \cC_\bp^{\alf\beta} + \cH_\bp^{\alf\beta}$, with
$\cE_\bp^{\alf\beta}$, $\cC_\bp^{\alf\beta}$, and $\cH_\bp^{\alf\beta}$ defined analogously.

Compared to other incommensurate magnetic states, the spiral state has the distinctive feature that the bare band is split in only two (not more) subbands, which is a consequence of a residual symmetry under translation combined with a spin rotation.
We also note that the global homogeneous magnetization resulting from the Zeeman coupling of the electron spin to an external magnetic field merely yields a shift of $h_\bp$ by a constant, which has no substantial effect on the results for the fields presently achieved in experiment.

%%%%%%%%%%%%%%%%%%%%%%%%%%%%%%%%%%%%%%%%%%%%%%%%%%%%%%%%%%%%%%%%%%%%%%%%%%%%%%%%%%

\subsection{Ordinary conductivity}

Expanding the grand canonical potential $\Om[\bA]$, Eq.~(\ref{OmA}), to second order in $\bA$, and performing the functional derivative with respect to the vector potential yields the current to linear order in $\bA$, from which we can read off the response kernel $K_{iq_0}^{\alf\beta}$. Using a basis in which the electron propagator is diagonal, one obtains \cite{voruganti92}
\begin{equation} \label{K1}
 K_{iq_0}^{\alf\beta} = e^2 \frac{T}{L} \sum_p {\rm tr} \big(
 \cG_{\bp,ip_0+iq_0} \tilde\lam_{\bp\bp}^\alf \cG_{\bp,ip_0} \tilde\lam_{\bp\bp}^\beta +
 \cG_{\bp,ip_0} \tilde\lam_{\bp\bp}^{\alf\beta} \big) \, .
\end{equation}
The first term, known as {\em paramagnetic} contribution, is due to the second-order term from Eq.~(\ref{OmA}) with first-order contributions for the vertices Eq.~(\ref{V_A}). The second term, known as {\em diamagnetic} contribution, arises from the first-order term in Eq.~(\ref{OmA}) with the second-order contribution for the vertex.
Using simple identities (see Appendix \ref{OrdinaryConductivity}), the response kernel can be rewritten as
\begin{eqnarray} \label{K2}
 K_{iq_0}^{\alf\beta} &=& e^2 \frac{T}{L} \sum_p {\rm tr} \big[
 (\cG_{\bp,ip_0+iq_0} - \cG_{\bp,ip_0}) \cE_\bp^\alf \cG_{\bp,ip_0} \cE_\bp^\beta
 \nonumber \\ && + \,
 (\cG_{\bp,ip_0+iq_0} - \cG_{\bp,ip_0}) \cF_\bp^\alf \cG_{\bp,ip_0} \cF_\bp^\beta
 \big] \, .
\end{eqnarray}
In this form all terms are quadratic in the propagators, and the property $K_{iq_0=0}^{\alf\beta} = 0$, which is dictated by gauge invariance, is manifestly satisfied.

Performing the analytic continuation to real frequencies and taking the limit $\om\to 0$ (see Appendix \ref{OrdinaryConductivity}) one obtains the following
expression for the dc conductivity
\begin{align} \label{sg}
 \sg^{\alf\beta} =&
 - e^2 \frac{\pi}{L} \sum_\bp \int \! d\eps \, f'(\eps) \left\{ 
 E_\bp^{+,\alf} E_\bp^{+,\beta} [A_\bp^+(\eps)]^2 \right. \nonumber \\ &\left.+ 
 E_\bp^{-,\alf} E_\bp^{-,\beta} [A_\bp^-(\eps)]^2 +
 2 F_\bp^\alf F_\bp^\beta A_\bp^+(\eps) A_\bp^-(\eps) \right\} \,,
\end{align}
where $f'(\eps)$ is the first derivative of the Fermi function
$f(\eps) = \big( e^{\eps/T} + 1 \big)^{-1}$, and
\begin{equation}
 A_\bp^\pm(\eps) = \frac{\Gam/\pi}{(\eps - E_\bp)^2 + \Gam^2}
\end{equation}
is the spectral function of the quasiparticles.\cite{footnote1}
The first two terms in Eq.~(\ref{sg}) are {\em intraband} contributions. They have the same form as for non interacting electrons with bare electron bands replaced by quasiparticle bands. The last term is an {\em interband} contribution involving states from both quasiparticle bands. For low temperatures and small $\Gam$ only momenta close to the quasiparticle Fermi surface, where $|E_\bp^+|$ or $|E_\bp^-|$ is small, contribute significantly to the conductivity.

The expression for the conductivity can be further simplified for small $\Gam$. The spectral functions $A^\pm(\eps)$ are Lorentzians of width $\Gam$.
Using $\pi \big[ A_\bp^\pm(\eps) \big]^2 \to (2\Gam)^{-1} \delta(\eps - E_\bp^\pm)$
for $\Gam \to 0$, one can simplify the intraband contribution to
\begin{equation}
 \sg_{\rm intra}^{\alf\beta} \to - e^2 \frac{\tau}{L} \sum_\bp \sum_{n=\pm}
 f'(E_\bp^n) \, E_\bp^{n,\alf} E_\bp^{n,\beta} \, ,
\end{equation}
where $\tau = 1/(2\Gam)$ is the relaxation time. This simplification holds when $\Gam$ is so small that the quasiparticle velocities $E_\bp^{\pm,\alf}$ are almost constant in a momentum range in which the variation of $E_\bp^\pm$ is of order $\Gam$.
Under the same assumption, and if in addition $\Gam \ll E_\bp^+ - E_\bp^-$, one has
\begin{equation}
 \pi A_\bp^+(\eps) A_\bp^-(\eps) \to \frac{\Gam}{(E_\bp^+ - E_\bp^-)^2} \,
 \big[ \delta(\eps - E_\bp^+) + \delta(\eps - E_\bp^-) \big] \, .
\end{equation}
The interband contribution to the conductivity can then be simplified to
\begin{equation}
 \sg_{\rm inter}^{\alf\beta} \to - e^2 \frac{\tau}{L} \sum_\bp \sum_{n=\pm}
 f'(E_\bp^n) \, \frac{F_\bp^\alf F_\bp^\beta}{\tau^2 (E_\bp^+ - E_\bp^-)^2} \, .
\end{equation}
The interband contribution is thus suppressed by a factor $\tau^{-2}$ compared to the intraband contribution for large $\tau$, and the naive formula for the conductivity, where bare bands are simply replaced by quasiparticle bands, can be applied.
$E_\bp^+ - E_\bp^- =
 2 \sqrt{\frac{1}{4}(\eps_{\bp+\bQ/2} - \eps_{\bp-\bQ/2})^2 + \Delta^2}$
is larger or equal to $2\Delta$, such that $\Gam \ll E_\bp^+ - E_\bp^-$ is satisfied for all $\bp$ if $\Gam \ll 2\Delta$. However, even for small $\Gam$, interband contributions may play a role near the transition between the paramagnetic and the antiferromagnetic phase, where $\Delta$ is also small.

For spiral states with wave vectors $\bQ$ for which one of the two components ($Q_x$ or $Q_y$) is $0$ or $\pi$, the off-diagonal components of the conductivity tensor vanish, that is, $\sg^{\alf\beta} = 0$ for $\alf \neq \beta$. Otherwise off-diagonal components are present. For the expressions obtained in the limit $\Gam \to 0$, this was already pointed out by Voruganti {\it et al.} \cite{voruganti92}

%%%%%%%%%%%%%%%%%%%%%%%%%%%%%%%%%%%%%%%%%%%%%%%%%%%%%%%%%%%%%%%%%%%%%%%%%%

\subsection{Hall conductivity}

Expanding the grand canonical potential $\Om[\bA]$ to third order yields the current to quadratic order in $\bA$. Inserting the decomposition $\bA = \ba^E + \ba^B$ and comparing with Eq.~(\ref{jq_exp}), one can read off the response kernel
\begin{align} \label{K_H}
 K_{\bq,iq_0}^{\alf\beta\gam} = \,& e^3 \frac{T}{L} \sum_p {\rm tr} \big(
 G_{\bp,ip_0} \lam_{\bp\bp}^{\alf\beta\gam}
 \nonumber \\
 +& G_{\bp^+,ip_0} \lam_{\bp^+\bp^-}^\gamma
     G_{\bp^-,ip_0} \lam_{\bp^-\bp^+}^{\alf\beta}
 \nonumber \\
 +& G_{\bp,ip_0+iq_0} \lam_{\bp\bp}^\beta
     G_{\bp,ip_0} \lam_{\bp\bp}^{\alf\gam}
 \nonumber \\
 +& G_{\bp^+,ip_0+iq_0} \lam_{\bp^+\bp^-}^\alf
     G_{\bp^-,ip_0} \lam_{\bp^-\bp^+}^{\beta\gam}
 \nonumber \\
 +& G_{\bp^+,ip_0+iq_0} \lam_{\bp^+\bp^-}^\alf
     G_{\bp^-,ip_0} \lam_{\bp^-\bp^+}^\gam
     G_{\bp^+,ip_0} \lam_{\bp^+\bp^+}^\beta 
 \nonumber \\
 +& G_{\bp^-,ip_0-iq_0} \lam_{\bp^-\bp^+}^\alf
     G_{\bp^+,ip_0} \lam_{\bp^+\bp^-}^\gam
     G_{\bp^-,ip_0} \lam_{\bp^-\bp^-}^\beta \big) \, ,
\end{align}
where $\bp^\pm=\bp\pm\frac{1}{2}\bq$.
Replacing $G$ by $\cG$ and $\lambda$ by $\tilde\lambda$ in the above equation, one can switch to the quasiparticle basis in which the electron propagator is diagonal.
$K_{\bq,iq_0}^{\alf\beta\gam}$ vanishes for $\bq = \b0$, as a consequence of gauge invariance. An explicit calculation confirming this property is presented in Appendix~\ref{HallConductivity}.
For small finite momenta, $K_{\bq,iq_0}^{\alf\beta\gam}$ can be expanded as
$K_{\bq,iq_0}^{\alf\beta\gam} = K_{iq_0}^{\alf\beta\gam\delta} q_\delta + \dots$.
To determine $\sg_H^{\alf\beta\gam}$ in a uniform magnetic field, it is sufficient to compute the first-order coefficient
\begin{equation}
 K_{iq_0}^{\alf\beta\gam\delta} = \frac{\partial}{\partial q_\delta}
 \left. K_{\bq,iq_0}^{\alf\beta\gam} \right|_{\bq=\b0} \, .
\end{equation}
From the six terms in Eq.~(\ref{K_H}), the first and the third do not contribute to $K_{iq_0}^{\alf\beta\gam\delta}$ since they are independent of $\bq$. The contribution from the second term, which is independent of $q_0$, also vanishes, as can be seen by a short explicit calculation. The evaluation of the remaining three contributions to $K_{iq_0}^{\alf\beta\gam\delta}$ is rather involved, since the matrix structure generates numerous terms. 
\begin{widetext}
After a lengthy calculation, which is presented in Appendix \ref{HallConductivity}, we obtain the comparatively simple result

\vspace{-0.3cm}
\begin{eqnarray} \label{Kabcd}
 K_{iq_0}^{\alf\beta\gam\delta} &=&
 - e^3 \frac{T}{4L} \sum_p {\rm tr} \big[
 (\cG_{\bp,ip_0+iq_0} - \cG_{\bp,ip_0-iq_0})
 \cE_\bp^\alf \cG_{\bp,ip_0} \cE_\bp^\delta \cG_{\bp,ip_0} \cE_\bp^{\beta\gam} \big]
 \nonumber \\
 && - e^3 \frac{T}{4L} \sum_p {\rm tr} \big[
 (\cG_{\bp,ip_0+iq_0} - \cG_{\bp,ip_0-iq_0})
 \cF_\bp^\alf \cG_{\bp,ip_0} \cE_\bp^\delta \cG_{\bp,ip_0} \cH_\bp^{\beta\gam} \big]
 \nonumber \\
 && - e^3 \frac{T}{2L} \sum_p {\rm tr} \big[
 (\cG_{\bp,ip_0+iq_0} - \cG_{\bp,ip_0-iq_0})
 \cF_\bp^\alf \cG_{\bp,ip_0} \cF_\bp^\delta \cG_{\bp,ip_0} \cE_\bp^{\beta\gam} \big]
 \nonumber \\
 && - (\alf \leftrightarrow \beta) - (\gam \leftrightarrow \delta) \, .
\end{eqnarray}
$K_{iq_0}^{\alf\beta\gam\delta}$ is antisymmetric under exchange of the first two indices ($\alf$ and $\beta$), as well as under exchange of the last two indices ($\gam$ and $\delta$). $K_{iq_0}^{\alf\beta\gam\delta}$ obviously vanishes for $q_0 = 0$.

Comparing Eq.~(\ref{Kabcd}) with Eq.~(\ref{Ksg_H}), and taking the limit $\om \to 0$, one obtains the following expression for the dc Hall conductivity:
\begin{equation} \label{sg_H}
 \sg_H^{\alf\beta\nu} =
 \sg_{H,{\rm intra}}^{\alf\beta\nu} + \sg_{H,{\rm inter}}^{\alf\beta\nu} \, ,
\end{equation}
with the intraband contribution
\begin{eqnarray}
 \sg_{H,{\rm intra}}^{\alf\beta\nu} &=&
 - e^3 \lim_{\om \to 0} \frac{\eps^{\nu\gam\delta}}{\om} \frac{T}{4L} \sum_{\bp,p_0} 
  {\rm tr} \big[
 (\cG_{\bp,ip_0+iq_0} - \cG_{\bp,ip_0-iq_0})
 \,
 \cE_\bp^\alf \cG_{\bp,ip_0} \cE_\bp^\delta \cG_{\bp,ip_0} \cE_\bp^{\beta\gam}
 - (\alf \leftrightarrow \beta) \big] \Big|_{iq_0 \to \om + i0^+} \, ,
\end{eqnarray}
and the interband contributions
\begin{eqnarray}
 \sg_{H,{\rm inter}}^{\alf\beta\nu} \hspace{-0.2cm}&=&
 - e^3 \lim_{\om \to 0} \frac{\eps^{\nu\gam\delta}}{\om} \frac{T}{4L} \sum_{\bp,p_0} 
  {\rm tr} \big[
 (\cG_{\bp,ip_0+iq_0} - \cG_{\bp,ip_0-iq_0})
 \,
 \cF_\bp^\alf \cG_{\bp,ip_0} \cE_\bp^\delta \cG_{\bp,ip_0} \cH_\bp^{\beta\gam}
 - (\alf \leftrightarrow \beta) \big] \Big|_{iq_0 \to \om + i0^+}
 \nonumber \\ &&
 - e^3 \lim_{\om \to 0} \frac{\eps^{\nu\gam\delta}}{\om} \frac{T}{2L} \sum_{\bp,p_0} 
  {\rm tr} \big[
 (\cG_{\bp,ip_0+iq_0} - \cG_{\bp,ip_0-iq_0})
 \,
 \cF_\bp^\alf \cG_{\bp,ip_0} \cF_\bp^\delta \cG_{\bp,ip_0} \cE_\bp^{\beta\gam}
 - (\alf \leftrightarrow \beta) \big] \Big|_{iq_0 \to \om + i0^+} \, .
\end{eqnarray}
%
% \begin{widetext}
% \end{widetext}
\end{widetext}

Already at this point one can see that $\sg_H^{\alf\beta\nu}$ vanishes for $\alf = \beta$, that is, the magnetic field does not affect the longitudinal conductivity to linear order in $\bB$.

The Matsubara sum and the analytic continuation to real frequencies, $iq_0 \to \om + i0^+$, can be performed analytically. All contributions to $\sg_H^{\alf\beta\nu}$ contain a Matsubara sum of the form
\begin{equation}
 K_{iq_0}^H = T \sum_{p_0} {\rm tr} \big[ (\cG_{ip_0+iq_0} - \cG_{ip_0-iq_0})
 \cM_1 \cG_{ip_0} \cM_2 \cG_{ip_0} \cM_3 \big] \, ,
\end{equation}
with arbitrary frequency-independent matrices $\cM_i$. Momentum dependencies are not written here. In Appendix \ref{HallConductivity} we show that
\begin{align} \label{hallcont}
 \lim_{\om \to 0} \frac{1}{\om} &K_{iq_0 \to \om + i0^+}^H  = 
 - \frac{2}{\pi} \int d\eps \, f(\eps)
 \nonumber \\
 \times& {\rm tr} \bigg[
 \Im\cG_\eps^A \cM_1 \partial_\eps \Re\Big(\cG_\eps^R \cM_2 \cG_\eps^R\Big) \cM_3 \nonumber\\&\hspace{0.5cm}+
 \Big(\partial_\eps \Re \cG_\eps^A\Big) \cM_1 \Im\Big(\cG_\eps^R \cM_2 \cG_\eps^R\Big) \cM_3 \bigg] ,
%  \hskip 7mm
\end{align}
where $\cG^A$ and $\cG^R$ are the advanced and retarded quasiparticle Green functions, respectively.

Using Eq.~(\ref{hallcont}), and performing a partial integration, the intraband contribution to the Hall conductivity can be written as
\begin{align} \label{sg_Hintra1}
 \sg_{H,{\rm intra}}^{\alf\beta\nu} &=
 -e^3 \pi^2 \frac{\eps^{\nu\gam\delta}}{3L}\sum_\bp
 \sum_{n=\pm}\,\int d\eps \,f'(\eps)\nonumber\\ &\times \big[E^{n,\alf}_\bp E^{n,\delta}_\bp E^{n,\beta\gam}_\bp- (\alf \leftrightarrow \beta)\big]
 \big[ A_\bp^n(\eps) \big]^3 \, .
\end{align}
For a two-dimensional electron system with a dispersion depending only on $p_x$ and $p_y$, and a perpendicular magnetic field (in the $z$ direction), the relevant component of the Hall conductivity reads
\begin{align} \label{sg_Hintra2}
 &\sg_{H,{\rm intra}}^{xyz} = 
 e^3 \frac{\pi^2}{3} \int \frac{d^2\bp}{(2\pi)^2} \sum_{n=\pm} \int d\eps \, f'(\eps)
 \nonumber \\
 \times& \big[ (E_\bp^{n,x})^2 E_\bp^{n,yy} - E_\bp^{n,x} E_\bp^{n,y} E_\bp^{n,xy}
 + (x \lra y) \big] \, \big[ A_\bp^n(\eps) \big]^3 \, .
\end{align}
Here and in the following $(x \lra y)$ denotes addition of the preceding terms with $x$ and $y$ exchanged. Note, however, that $\sg_H^{xyz}$ is antisymmetric in $x$ and $y$.
For small $\Gam$, the product of spectral functions can be replaced by a Dirac delta function,
\begin{equation}
 \big[ A^\pm(\eps) \big]^3 \to \frac{3}{8\pi^2} \Gam^{-2} \delta(\eps-E^\pm_\bp) ,
\end{equation}
so that the integral over $\eps$ can be performed, yielding
\begin{align} \label{sg_Hintra3}
 \sg_{H,{\rm intra}}^{xyz} \rightarrow&
 \frac{e^3\tau^2}{2} \int \frac{d^2\bp}{(2\pi)^2} \sum_{n=\pm} f'(E^n_\bp) \nonumber \\ & \times 
 \big[ (E_\bp^{n,x})^2 E_\bp^{n,yy} - E_\bp^{n,x} E_\bp^{n,y} E_\bp^{n,xy}
 + (x \lra y) \big] \, ,
\end{align}
where $\tau=1/(2\Gam)$. 
Using $f'(E_\bp^2) \, E_\bp^{n,\alf} = \partial_{p_\alf} f(E_\bp^n)$, and performing a partial integration, this can also be written as
\begin{align} \label{sg_Hintra4}
 \sg_{H,{\rm intra}}^{xyz} = 
 - e^3 \tau^2 &\int \frac{d^2\bp}{(2\pi)^2} \sum_{n=\pm} f(E_\bp^n) \nonumber\\&\times\big[ 
 E_\bp^{n,xx} E_\bp^{n,yy} - E_\bp^{n,xy} E_\bp^{n,yx} \big] \, .
\end{align}
Equations (\ref{sg_Hintra3}) and (\ref{sg_Hintra4}) agree with the corresponding expressions  derived by Voruganti {\it et al.}\cite{voruganti92}

Applying Eq.~(\ref{hallcont}) to the interband contribution, one obtains
\begin{widetext}
%
% \vspace{-0.5cm}
\begin{align} \label{sg_Hinter1}
 \sg_{H,{\rm inter}}^{\alf\beta\nu} =
 - \,\, e^3 \pi^2 \frac{\eps^{\nu\gam\delta}}{L} \sum_\bp \sum_{n=\pm} &\int d\eps \,
 f'(\eps) \, \big[ A_\bp^n(\eps) \big]^2 A_\bp^{-n}(\eps) \,
 \big(F^\alf_\bp H^{\beta\gam}_\bp-F^\beta_\bp H^{\alf\gam}_\bp\big)
 E^{n,\delta}_\bp \nonumber \\
 + 2 e^3 \pi^2 \frac{\eps^{\nu\gam\delta}}{L} \sum_\bp \sum_{n=\pm} &\int d\eps \,
 f(\eps) \, A_\bp^+(\eps) A_\bp^-(\eps) \,
 \frac{A_\bp^+(\eps) - A_\bp^-(\eps)}{E_\bp^+ - E_\bp^-}
 \left[
 F^\alf_\bp
 \big( H^{\beta\gam}_\bp E^{n,\delta}_\bp + F_\bp^\gam E^{n,\beta\delta}_\bp \big) -
 ( \alf \lra \beta ) \right] \, .
\end{align}
Note that we have chosen a mixed representation with a Fermi function derivative in the first term and the Fermi function in the second. Performing a partial integration on the second term would lead to an integrand with contributions away from the quasiparticle energies, even for small $\Gam$.
Specifying again to two dimensions with a magnetic field in the $z$ direction, we get
\begin{align} \label{sg_Hinter2}
 \sg_{H,{\rm inter}}^{xyz} =
 - \,\,e^3 \pi^2 &\int \frac{d^2\bp}{(2\pi)^2} \sum_{n=\pm} \int\! d\eps \,
 f'(\eps) \big[ A_\bp^n(\eps) \big]^2 A_\bp^{-n}(\eps) \,
 \big[ F^x_\bp \big( H^{yx}_\bp E^{n,y}_\bp - H^{yy}_\bp E^{n,x}_\bp \big) +
 (x \lra y) \big] \nonumber \\
 + 2 e^3 \pi^2 &\int \frac{d^2\bp}{(2\pi)^2} \sum_{n=\pm} \int\! d\eps \,
 f(\eps) \, A_\bp^+(\eps) A_\bp^-(\eps) \,
 \frac{A_\bp^+(\eps) - A_\bp^-(\eps)}{E_\bp^+ - E_\bp^-}
 \nonumber\\
 & \hspace{4cm}\times \left[
 F^x_\bp \big( H^{yx}_\bp E^{n,y}_\bp - H^{yy}_\bp E^{n,x}_\bp +
 F^x_\bp E^{n,yy}_\bp - F^y_\bp E^{n,yx}_\bp \big) + (x \lra y) \right] \, .
\end{align}
For small $\Gam$, the products of spectral functions can be replaced by delta functions,
\begin{equation}
 \big[ A_\bp^n(\eps) \big]^2 A_\bp^{-n}(\eps) \to
 \frac{\delta(\eps - E_\bp^n)}{2\pi^2 (E_\bp^+ - E_\bp^-)^2} \, ,
\end{equation}
so that the $\eps$ integral can be performed, yielding
\begin{eqnarray} \label{sg_Hinter3}
 \sg_{H,{\rm inter}}^{xyz} =
 &-& \, \frac{e^3}{2} \!\int\! \frac{d^2\bp}{(2\pi)^2} \sum_{n=\pm}
 \frac{f'(E_\bp^n)}{(E_\bp^+ - E_\bp^-)^2} \,
 \big[ F^x_\bp \big( H^{yx}_\bp E^{n,y}_\bp - H^{yy}_\bp E^{n,x}_\bp \big) +
 (x \lra y) \big] \nonumber \\
 &+& e^3 \!\int\! \frac{d^2\bp}{(2\pi)^2} \sum_{n=\pm}
 \frac{f(E_\bp^+) - f(E_\bp^-)}{(E_\bp^+ - E_\bp^-)^3}
 \left[
 F^x_\bp \big( H^{yx}_\bp E^{n,y}_\bp - H^{yy}_\bp E^{n,x}_\bp +
 F^x_\bp E^{n,yy}_\bp - F^y_\bp E^{n,yx}_\bp \big) + (x \lra y) \right] .
\end{eqnarray}
\end{widetext}

The above simplifications for $\Gam \to 0$ apply under the same conditions as for the ordinary conductivity, that is, the momentum-dependent functions $E_\bp^\alf$, $F_\bp^\alf$, etc., must be almost constant in the momentum range in which the variation of $E_\bp^\pm$ is of order $\Gam$, and $\Gam$ must be much smaller than $E_\bp^+ - E_\bp^-$. Here, for the Hall conductivity, the interband contributions are also suppressed by a factor $\tau^{-2}$ compared to the intraband contributions.

We finally emphasize that our derivation is valid under the assumption of a momentum-independent magnetic gap $\Delta$ and a momentum-independent relaxation rate $\Gam$.
A generalization to a momentum-dependent gap and relaxation rate is not straightforward, since numerous additional terms appear.

%%% Results %%%%%%%%%%%%%%%%%%%%%%%%%%%%%%%%%%%%%%%%%%%%%%%%%%%%%%%%%%%%%%

\section{Results}

We now present and discuss results for the longitudinal conductivity and the Hall conductivity. Motivated by the recent charge transport experiments in cuprates, \cite{badoux16,collignon17} we compute these quantities with a phenomenological ansatz for a doping dependent magnetic gap $\Delta(p)$, in close analogy to previous theoretical studies.\cite{storey16,eberlein16,chatterjee17}
In particular, we will study the size of interband contributions to the conductivities, which were neglected in earlier calculations for N\'eel and spiral magnetic states.
Interband contributions have been taken into account, however, in a calculation of the optical conductivity in a $d$-density wave state, \cite{aristov05} and in a very recent evaluation of the longitudinal dc conductivity in the spiral state. \cite{chatterjee17}

We choose a tight-binding band structure
\begin{align}
 \eps_\bp = &-2t (\cos p_x + \cos p_y) - 4t' \cos p_x \cos p_y \nonumber \\
 &- 2t''[\cos(2p_x) + \cos(2p_y)] \, ,
\end{align}
where $t$, $t'$, and $t''$ are nearest-, second-nearest-, and third-nearest-neighbor hopping amplitudes, respectively, on a square lattice with lattice constant $a=1$.
We choose $t$ as our unit of energy. Hopping amplitudes in cuprates have been determined by downfolding {\it ab initio} band structures on effective single-band Hamiltonians.\cite{andersen95,pavarini01} The ratio $t'/t$ is negative in all cuprate superconductors, ranging from $-0.15$ in LSCO to $-0.35$ in $\text{Bi}_2\text{Sr}_2\text{CaCu}_2\text{O}_{8+\delta}$.

Theoretical results for spiral states in the two-dimensional $t$-$J$ model \cite{sushkov04} suggest a linear doping dependence of the magnetic gap of the form
\begin{equation}
 \Delta(p) = D (p^* - p) \Theta(p^* - p) \, ,
\end{equation}
where $D$ is a prefactor, and $p^*$ is the critical doping at which the magnetic order vanishes.
Both $D$ and $p^*$ are material dependent and need to be fitted to experimental data.
A linear doping dependence of the gap for $p < p^*$ is also found in resonating valence bond mean-field theory for the $t$-$J$ model.\cite{yang06}
The pseudogap temperature scale $T^*$ seen in experiments also vanishes linearly at a certain critical doping $p^*$.
In Ref.~\onlinecite{eberlein16}, a quadratic doping dependence of $\Delta(p)$ was considered, too.

The wave vector of the incommensurate magnetic states obtained in the theoretical literature\cite{schulz90,kato90,fresard91,raczkowski06,igoshev10,chubukov92,chubukov95,metzner12,yamase16,zheng17,vilardi18,shraiman89,kotov04}
has the form $\bQ = (\pi-2\pi\eta,\pi)$, or symmetry related, that is,
$(-\pi+2\pi\eta,\pi)$, $(\pi,\pi-2\pi\eta)$, and $(\pi,-\pi+2\pi\eta)$.
Here $\eta > 0$, the so-called incommensurability, measures the deviation from the N\'eel wave vector $(\pi,\pi)$. Peaks in the magnetic structure factors seen in neutron-scattering experiments are also situated at such wave vectors.\cite{yamada98,haug09,haug10}
The incommensurability $\eta$ is a monotonically increasing function of doping. In Ref.~\onlinecite{eberlein16}, the doping dependence of $\eta$ was determined by minimizing the mean-field free energy, resulting in $\eta \approx p$, which is roughly consistent with experimental observations in LSCO. \cite{yamada98} In YBCO $\eta$ values below $p$ are observed, \cite{haug10} and functional renormalization group calculations for the Hubbard model also yield $\eta < p$. \cite{yamase16}
The precise doping dependence of the incommensurability has no significant effect on the transport properties. Hence, we simply choose $\eta = p$ in most of our results, but show a comparison to results obtained with other choices of $\eta(p)$ in our final fit of the Hall number to experimental data.

For small relaxation rates $\Gam$, interband contributions are suppressed by a factor $\Gam^2$ compared to intraband contributions to the conductivities (see Sec.~II).
To get a feeling for the typical size of $\Gam$ in the recent high-field experiments, we estimate $\Gam$ from the experimental result $\om_c \tau = 0.075$ reported for Nd-LSCO samples at zero temperature by Collignon {\it et al.}\cite{collignon17}
The cyclotron frequency can be written as $\om_c = \frac{|e|B}{m_c}$, which defines the cyclotron mass $m_c$. For free electrons $m_c$ is just the bare electron mass $m_e$. Inserting the applied magnetic field of 37.5 T and assuming $m_c = m_e$, one obtains $\Gam = (2\tau)^{-1} \approx 0.03 \, \text{eV}$. With the typical value $t \approx 0.3 \, \text{eV}$ for the nearest neighbor hopping amplitude in cuprates, one thus gets $\Gam/t \approx 0.1$. The cyclotron mass in cuprates is actually larger than the bare electron mass. Mass ratios $m_c/m_e$ equal to 3 or even larger have been observed. \cite{stanescu08} Hence, $\Gam/t = 0.1$ is just an upper bound; the actual value can be expected to be even smaller. Indeed, an estimate from the observed residual resistivity in Nd-LSCO yields $\Gam \approx 0.008 \, \text{eV}$. \cite{taillefer18}

The relaxation rates in cuprate superconductors are actually momentum dependent. However, we do not expect the momentum dependence to affect the order of magnitude of interband contributions. Concerning the doping dependence of $\Gam$, we are using experimental input. Magnetoresistance data suggest that the electron mobility does not change significantly in the doping range where the Hall number drop is observed.\cite{collignon17} Since the mobility is directly proportional to the inverse relaxation rate, we therefore choose $\Gam$ independent of doping.

%%%%%%%%%%%%%%%%%%%%%%%%%%%%%%%%%%%%%%%%%%%%%%%%%%%%%%%%%%%%%%%%%%%%%%%%%%%%%%%%%%%%

\subsection{Longitudinal conductivity}

\begin{figure}
\centering
\includegraphics[width=7cm]{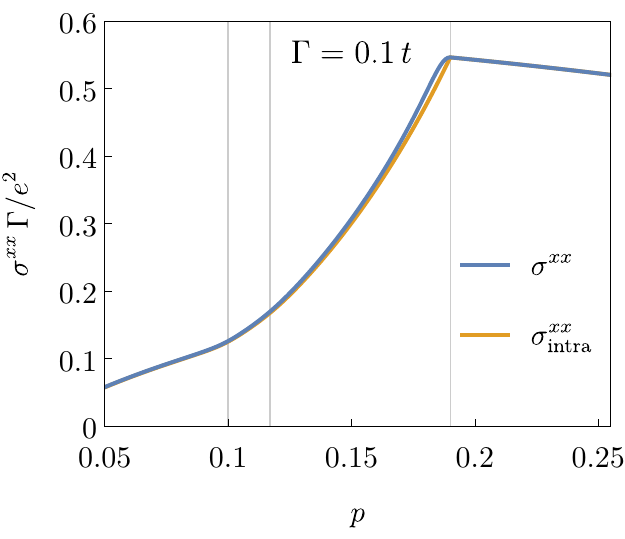}
\includegraphics[width=7cm]{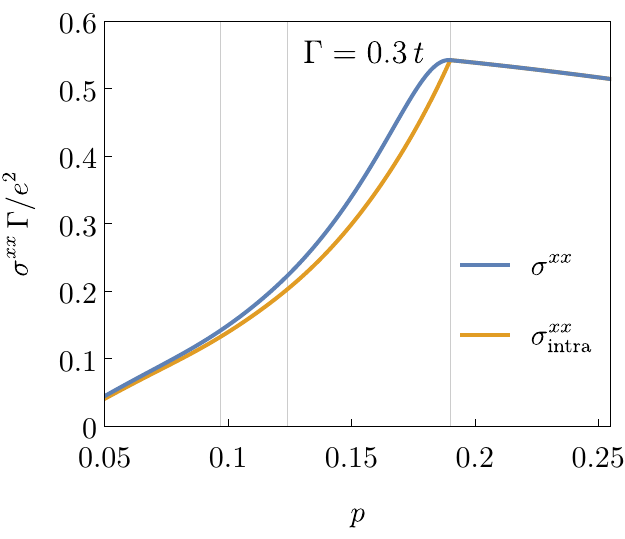}
\caption{Longitudinal conductivity $\sg^{xx}$ at zero temperature as a function of doping $p$ for a doping dependent magnetic order parameter $\Delta(p) = 12t (p^* - p) \Theta(p^* - p)$ with $p^* = 0.19$. The intraband contribution $\sg_{\rm intra}^{xx}$ is also shown for comparison. The hopping parameters are $t'/t = -0.3$ and $t''/t = 0.2$. Top: $\Gam/t = 0.1$. Bottom: $\Gam/t = 0.3$. The vertical lines indicate changes of the Fermi-surface topology at the three doping values $p_e^*$, $p_h^*$, and $p^*$.}
\label{fig_cond}
\end{figure}

In Fig.~\ref{fig_cond} we show results for $\sg^{xx}$ as obtained from Eq.~(\ref{sg}) at zero temperature for two values of the relaxation rate $\Gam$. For the hopping parameters we chose values used for YBCO in the literature. The critical doping $p^* = 0.19$ is the onset doping for the Hall number drop observed in the experiments on YBCO by Badoux {\it et al.} \cite{badoux16} The total conductivity $\sg^{xx}$ is compared to the intraband contribution $\sg_{\rm intra}^{xx}$.
One can see a pronounced drop of the conductivity for $p < p^*$, as expected from the drop of charge-carrier density in the spiral state.
For $\Gam/t = 0.1$ the interband contributions are practically negligible, while for $\Gam/t = 0.3$ they are already sizable. In particular, the interband contributions shift the drop of $\sg^{xx}$ induced by the spiral order toward smaller values of $p$, and they smooth the sharp kink exhibited by $\sg^{xx}$ at $p^*$ for $\Gam \to 0$.

Chatterjee {\it et al.} \cite{chatterjee17} have derived expressions for the electrical and the heat conductivities in the spiral state, for a momentum independent relaxation rate, and showed that the two quantities are related by the Wiedemann-Franz law. While their formulas for the conductivities have a different form than ours, we have checked that the numerical results are consistent.
The spiral state exhibits a pronounced nematicity in the longitudinal conductivity.

\begin{figure}
\centering
\includegraphics[width=7cm]{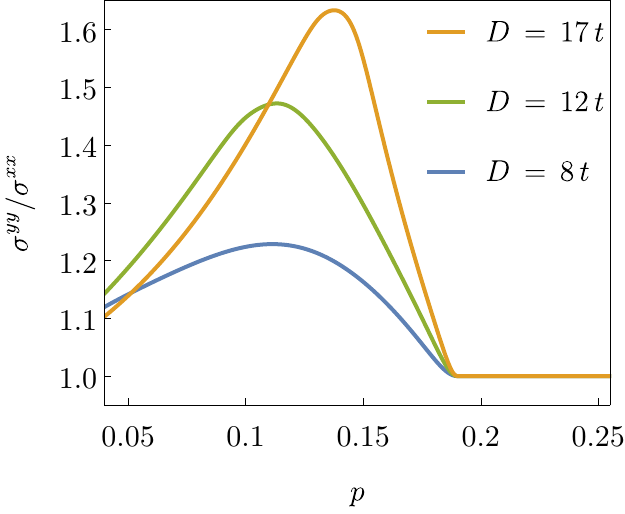}
\caption{Anisotropy ratio of the longitudinal conductivity $\sg^{yy}/\sg^{xx}$ at zero temperature as a function of doping $p$ for three choices of $D$. The band parameters are the same as in Fig.~\ref{fig_cond}, and the relaxation rate is $\Gam/t = 0.1$.}
\label{fig_anisotropy}
\end{figure}

For $\bQ = (\pi - 2\pi\eta,\pi)$, with an incommensurability in the $x$ direction, the conductivity in the $y$ direction is larger than in the $x$ direction.
In Fig.~\ref{fig_anisotropy} we show the ratio $\sg^{yy}/\sg^{xx}$ as a function of doping for the same band parameters as in Fig.~\ref{fig_cond} and various choices for $D$, with $\Gam/t = 0.1$. The anisotropy increases smoothly upon lowering the doping from the critical point $p^*$, and it decreases upon approaching half filling, where $\eta$ vanishes such that the square lattice symmetry is restored.
A pronounced temperature and doping dependent in-plane anisotropy of the longitudinal conductivities with conductivity ratios up to 2.5 has been observed in YBCO by Ando {\it et al.} \cite{ando02}
There is no contribution to $\sg^{xy}$ and $\sg^{yx}$ for spiral states with a wave vector of the form $\bQ = (\pi - 2\pi\eta,\pi)$.

\begin{figure}
\centering %5.69
\includegraphics[width=6.7cm]{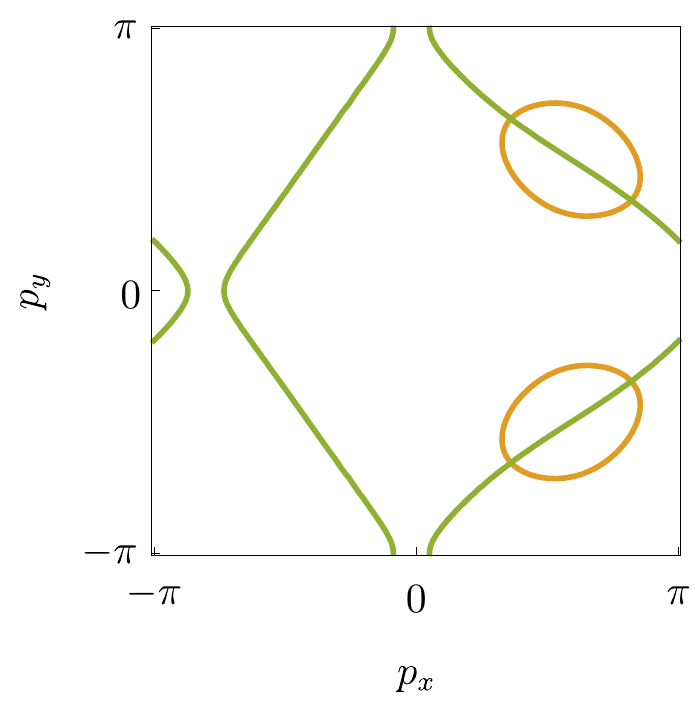}  
\includegraphics[width=6.7cm]{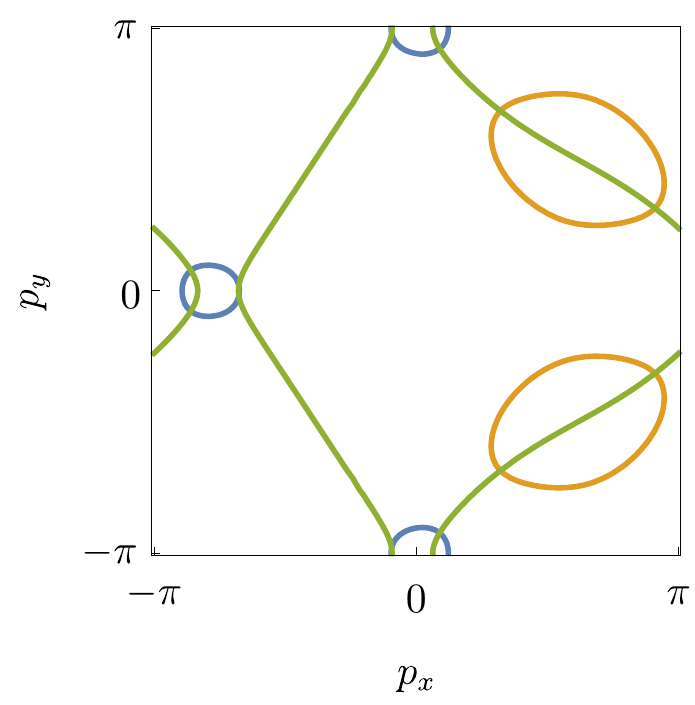}  
\includegraphics[width=6.7cm]{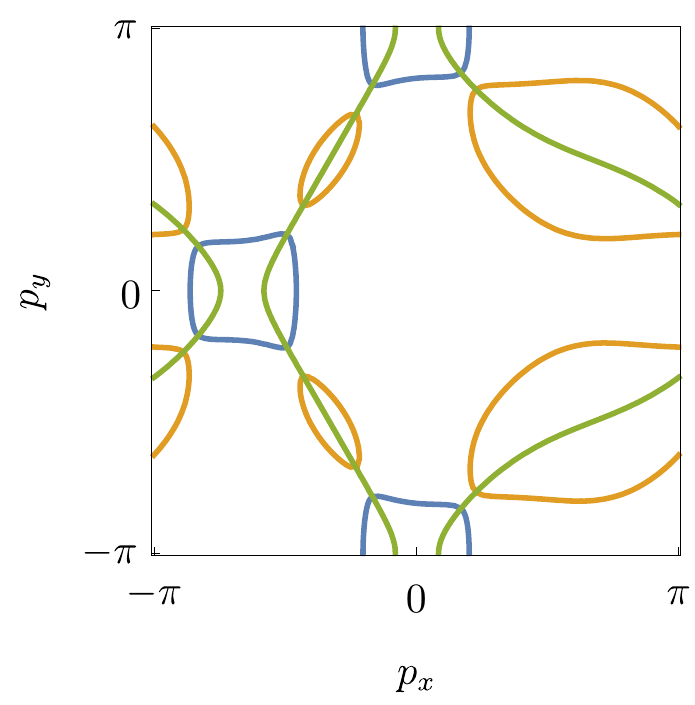}
\caption{Quasiparticle Fermi surfaces for $p = 0.09$, $p = 0.115$, and $p = 0.17$ (from top to bottom). Fermi-surface sheets surrounding hole (orange) and electron (blue) pockets correspond to zeros of $E_{\bp+\bQ/2}^-$ and $E_{\bp+\bQ/2}^+$, respectively. The green ``nesting'' line indicates momenta $\bp$ satisfying the condition $\eps_{\bp} = \eps_{\bp+\bQ}$. The band and gap parameters are the same as in Fig.~\ref{fig_cond}, with $\Gam/t = 0.1$.}
\label{fig_fs}
\end{figure}

For $p < p_e^*$, the quasiparticle Fermi surface consists exclusively of hole pockets, while for $p_e^* < p < p^*$ also electron pockets are present.
Note that $p_e^*$ depends (slightly) on the relaxation rate $\Gam$, since the relation between the chemical potential $\mu$ and the density depends on $\Gam$.
For $p < p_h^*$ there are only two hole pockets, while for $p_h^* < p < p^*$ a second (smaller) pair of hole pockets appear.
In Fig.~\ref{fig_fs} we plot the quasiparticle Fermi surfaces for three choices of the doping $p$: for $p < p_e^*$, for $p_e^* < p < p_h^*$, and for $p_h^* < p < p^*$. For a more transparent representation of the Fermi-surface topology, we shift the momentum by $\bQ/2$ and plot zeros of $E_{\bp+\bQ/2}^\pm$ instead of zeros of $E_\bp^\pm$. The Fermi surfaces look thus similar to those shown in Ref.~\onlinecite{eberlein16}, where the shift by $\bQ/2$ was already included in the definition of the quasiparticle energies.
At $p = p^*$ electron and hole pockets merge, and for $p > p^*$ there is only a single large Fermi-surface sheet, which is closed around the unoccupied (hole) states. The doping dependence of the conductivity changes its slope at $p_e^*$, while there is no pronounced feature at $p_h^*$. However, choosing a smaller relaxation rate $\Gam \ll 0.1t$, a change of slope of $\sg^{xx}$ is visible also at $p_h^*$, while no pronounced feature of $\sg^{yy}$ is visible.
The sequence of Fermi-surface topologies as a function of doping depends on the doping dependence of $\eta$. The above results were obtained for $\eta = p$. Choosing, for example, a smaller $\eta(p)$, one may have four (not just two) hole pockets at low doping.

It is instructive to see which quasiparticle states yield the dominant contributions to the conductivity. In two dimensions, the conductivity in Eq.~(\ref{sg}) is given by a momentum integral of the form
$\sg^{\alf\beta} = \int \frac{d^2\bp}{(2\pi)^2} \, \sg^{\alf\beta}(\bp)$.
The Fermi function derivative $f'(\eps)$ restricts the energies $\eps$ up to values of order $T$. For $T=0$, one has $f'(\eps) = - \delta(\eps)$. For small $\Gam$ the quasiparticle spectral functions $A_\bp^{\pm}(\eps)$ are peaked at the quasiparticle energies. Hence, for low $T$ and small or moderate $\Gam$, the dominant contributions to the conductivity come from momenta where either $|E_\bp^+|$ or $|E_\bp^-|$ is small, that is, in particular from momenta near the quasiparticle Fermi surfaces.

\begin{figure*}
\centering
\includegraphics[width=8cm]{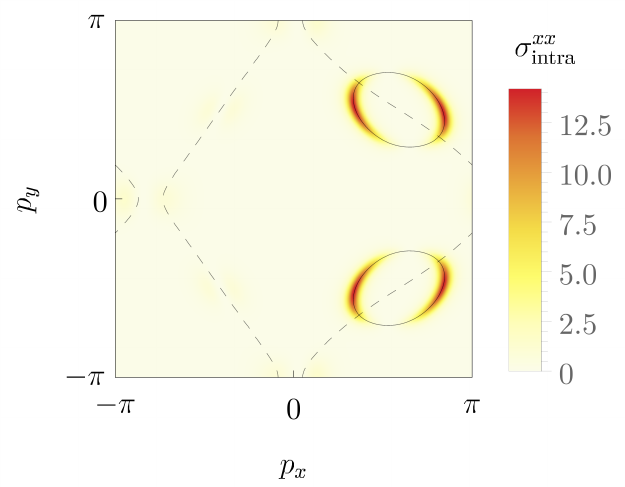}
\includegraphics[width=8cm]{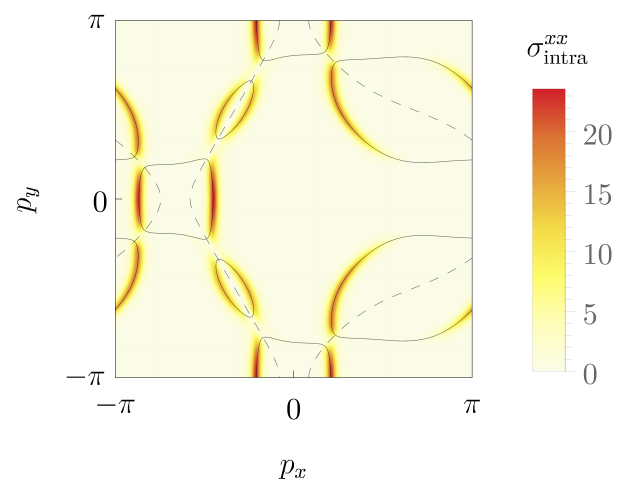} \\
\includegraphics[width=8cm]{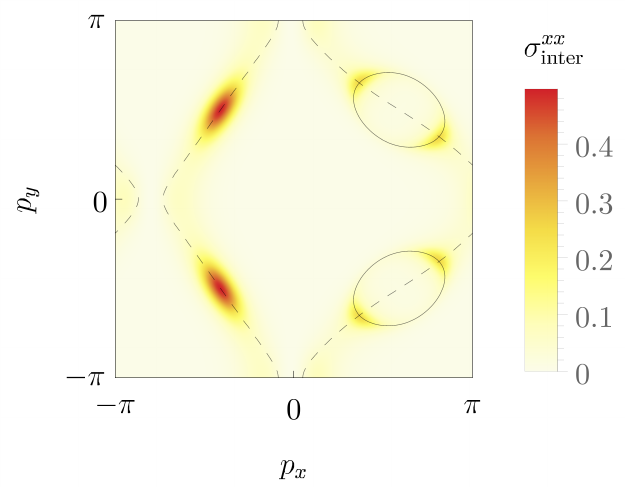}
\includegraphics[width=8cm]{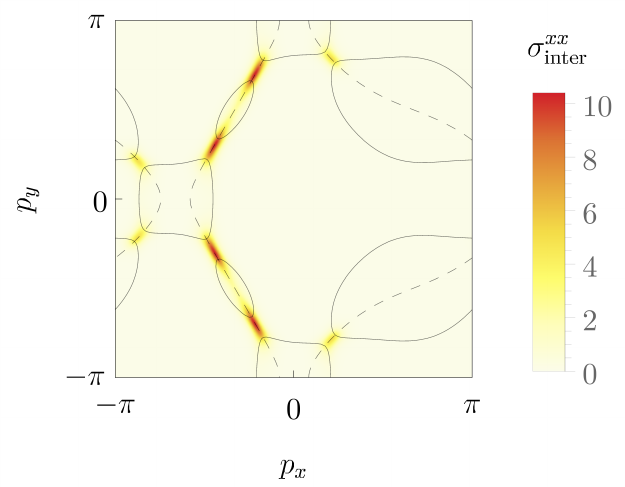}
\caption{Top: Color plot of the momentum resolved intraband contribution to the longitudinal conductivity $\sg_{\rm intra}^{xx}(\bp+\bQ/2)$ for $p = 0.09$ (left) and $p = 0.17$ (right). Bottom: Interband contribution $\sg_{\rm inter}^{xx}(\bp+\bQ/2)$ for the same choices of $p$. The band and gap parameters are the same as in Fig.~\ref{fig_cond}, and the relaxation rate is $\Gam/t = 0.3$. The Fermi surfaces and the nesting line (cf.\ Fig.~\ref{fig_fs}) are plotted as thin black lines.}
\label{fig_sgp}
\end{figure*}

In Fig.~\ref{fig_sgp} we show color plots of $\sg_{\rm intra}^{xx}(\bp+\bQ/2)$ and $\sg_{\rm inter}^{xx}(\bp+\bQ/2)$ in the Brillouin zone. Although a sizable $\Gam/t = 0.3$ has been chosen, the intraband contributions are clearly restricted to the vicinity of the quasiparticle Fermi surface. Variations of the size of intraband contributions along the Fermi surfaces are due to the momentum dependence of the quasiparticle velocities $E_\bp^{\pm,x} = \partial E_\bp^\pm/\partial p_x$.
The interband contributions are particularly large near the ``nesting line'' defined by $\eps_{\bp+\bQ} = \eps_{\bp}$, where the direct band gap between the quasiparticle energies $E_\bp^+$ and $E_\bp^-$ assumes the minimal value $2\Delta$. For $p=0.09$, the largest interband contributions come from regions on the nesting line remote from the Fermi surfaces. Note, however, that they are much smaller than the intraband contributions, and $|E_\bp^-|$ has a local minimum in these regions. For $p=0.17$, the interband contributions are generally larger, and they are concentrated in regions between neighboring electron and hole pockets.

%%%%%%%%%%%%%%%%%%%%%%%%%%%%%%%%%%%%%%%%%%%%%%%%%%%%%%%%%%%%%%%%%%%%%%%%%%%%%%%%%%%%

\subsection{Hall conductivity}

The Hall conductivity $\sg_H^{xyz}$ and the longitudinal conductivities determine the {\em Hall coefficient}
\begin{equation}
 R_H = \frac{\sg_H^{xyz}}{\sg^{xx} \sg^{yy}} \, .
\end{equation}
Unlike the longitudinal and Hall conductivities, the Hall coefficient is finite in the limit $\Gam \to 0$.
In the independent electron approximation, there are special cases where the Hall coefficient is determined by the charge density $\rho_c$ via the simple relation $R_H = \rho_c^{-1}$. For free electrons with a parabolic dispersion, this relation holds for any magnetic field, with $\rho_c = e n_e$. For band electrons it still holds in the high-field limit $\om_c \tau \gg 1$, if the semiclassical electron orbits of all occupied (or all unoccupied) states are closed.\cite{ashcroft76} For Fermi surfaces enclosing unoccupied states, the relevant charge density is then $\rho_c = |e| n_h$, where $n_h$ is the density of holes. If both electron and holelike Fermi surfaces are present, one has $\rho_c = e (n_e - n_h)$. \cite{ashcroft76}
Results for the Hall conductivity are thus frequently represented in terms of the so-called {\em Hall number} $n_H$, defined via the relation
\begin{equation}
 R_H = \frac{1}{|e| \, n_H} \, .
\end{equation}
However, $n_H$ is given by the electron and hole densities only in the special cases described above. In particular, in the low-field limit $\om_c \tau \ll 1$ which applies to the recent ``high'' magnetic field experiments for cuprates, there is no guarantee that $n_H$ is equal to a charge-carrier density.

\begin{figure*}
\centering
\includegraphics[width=7cm]{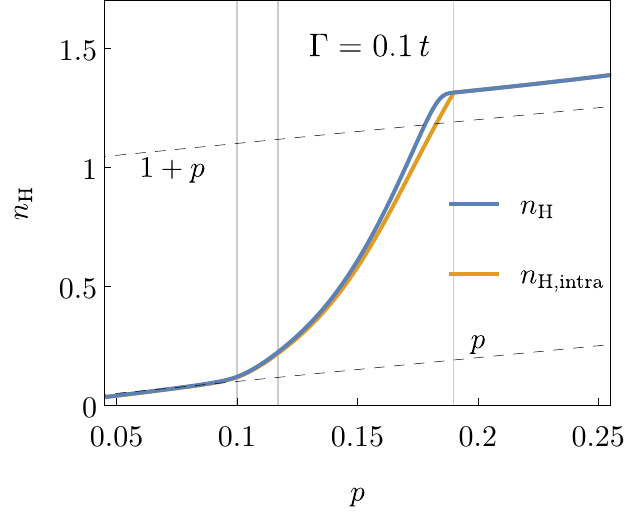}
\includegraphics[width=7cm]{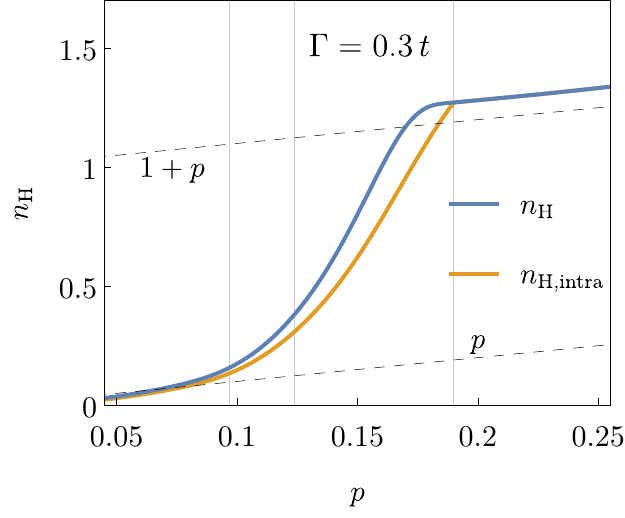}
\caption{Hall number $n_H$ as a function of doping $p$ for a doping dependent magnetic order parameter $\Delta(p) = 12t (p^* - p) \Theta(p^* - p)$ with $p^* = 0.19$. The intraband contribution $n_{H,{\rm intra}}$ is also shown for comparison. The straight dashed lines correspond to the naive expectation for large and reconstructed Fermi surfaces, $n_H = 1+p$ and $n_H = p$, respectively. The vertical lines indicate the three special doping values $p_e^*$, $p_h^*$, and $p^*$. The hopping parameters are $t'/t = -0.3$ and $t''/t = 0.2$. Left: $\Gam/t = 0.1$. Right: $\Gam/t = 0.3$.}
\label{fig_hall}
\end{figure*}

In Fig.~\ref{fig_hall} we show results for the Hall number as obtained from the Hall conductivity in Eqs.~(\ref{sg_Hintra2}) and Eq.~(\ref{sg_Hinter2}).
The hopping and gap parameters are the same as in Fig.~\ref{fig_cond}.
The Hall number obtained from the total Hall conductivity is compared to the one obtained by neglecting interband contributions, that is, taking only $\sg_{H,{\rm intra}}^{xyz}$ and $\sg_{\rm intra}^{\alf\alf}$ into account.
For $p \geq p^*$, where $\Delta = 0$, the Hall number is slightly above the value $1+p$ corresponding to the density of holes enclosed by the (large) Fermi surface. This is also seen in experiment in YBCO \cite{badoux16} and Nd-LSCO. \cite{collignon17} Note that $n_H$ is not expected to be equal to $1+p$ for $\om_c \tau \ll 1$, since the dispersion $\eps_\bp$ is not parabolic.
For $p < p^*$ the Hall number drops drastically. For $\Gam/t = 0.1$ the interband contributions are again quite small, as already observed for the longitudinal conductivity, and the Hall number gradually approaches the value $p$ upon lowering $p$. Hence, the naive expectation that the Hall number is given by the density of holes in the hole pockets turns out to be correct for sufficiently small $p$. Visible deviations from $n_H = p$ set in for $p > p_e^*$, where the electron pockets emerge. For $\Gam/t = 0.3$ interband contributions are sizable. They shift the onset of the drop of $n_H$ to smaller doping.

\begin{figure*}
\centering
\includegraphics[width=8cm]{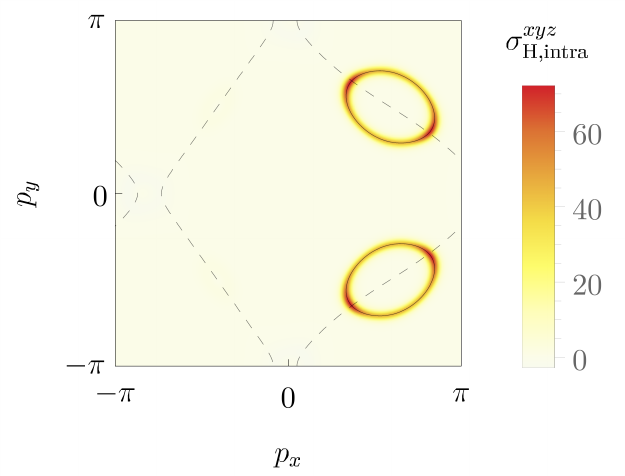}
\includegraphics[width=8cm]{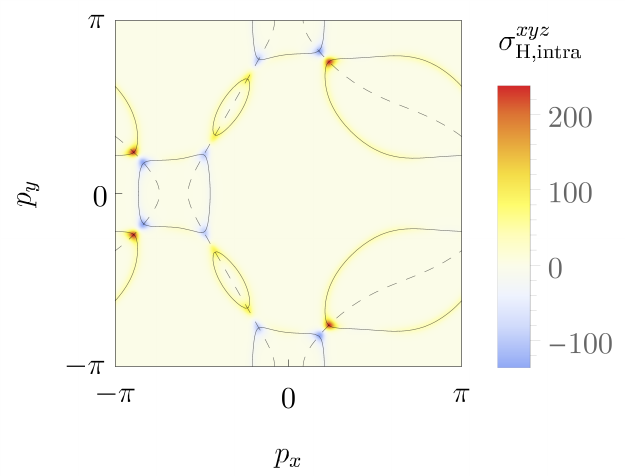} \\
\includegraphics[width=8cm]{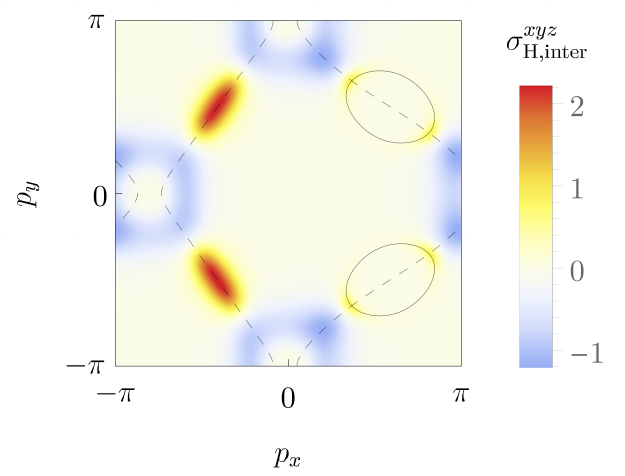}
\includegraphics[width=8cm]{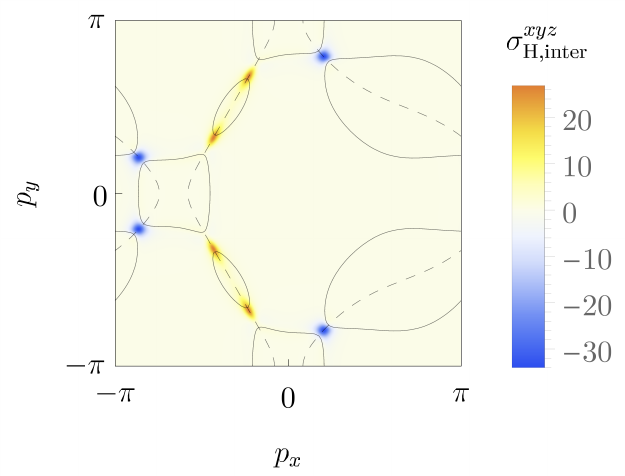}
\caption{Top: Color plot of the momentum resolved intraband contribution to the Hall conductivity $\sg_{H,{\rm intra}}^{xyz}(\bp+\bQ/2)$ for $p = 0.09$ (left) and $p = 0.17$ (right). Bottom: Interband contribution $\sg_{H,{\rm inter}}^{xyz}(\bp+\bQ/2)$ for the same choices of $p$. The band and gap parameters are the same as in Fig.~\ref{fig_hall}, and the relaxation rate is $\Gam/t = 0.3$. The Fermi surfaces and the nesting line (cf.\ Fig.~\ref{fig_fs}) are plotted as thin black lines.}
\label{fig_sgHp}
\end{figure*}
The Hall conductivity in Eqs.~(\ref{sg_Hintra2}) and (\ref{sg_Hinter2}) is given by a momentum integral of the form  
$\sg_H^{xyz} = \int \frac{d^2\bp}{(2\pi)^2} \sg_H^{xyz}(\bp)$.
To see which momenta, that is, which quasiparticle states, contribute most significantly to the Hall conductivity, we show color plots of
$\sg_{H,{\rm intra}}^{xyz}(\bp+\bQ/2)$ and $\sg_{H,{\rm inter}}^{xyz}(\bp+\bQ/2)$ for two choices of the hole doping; see Fig.~\ref{fig_sgHp}.
The intraband contributions are concentrated near the quasiparticle Fermi surfaces, due to the peaks in $f'(\eps)$ and in the spectral functions, as for the longitudinal conductivity. Contributions from hole pockets count positively, and those from electron pockets negatively, as expected. The contributions are particularly large near crossing points of the Fermi surfaces with the nesting line, where the Fermi surfaces have a large curvature. The interband contributions lie mostly near the nesting line, not necessarily close to Fermi surfaces. For $p=0.17$ they are concentrated in small regions between electron and hole pockets.

%%%%%%%%%%%%%%%%%%%%%%%%%%%%%%%%%%%%%%%%%%%%%%%%%%%%%%%%%%%%%%%%%%%%%%%%%%%%%%%%%%%

\subsection{Comparison to experiments}

\begin{figure}
\centering
\includegraphics[width=7.5cm]{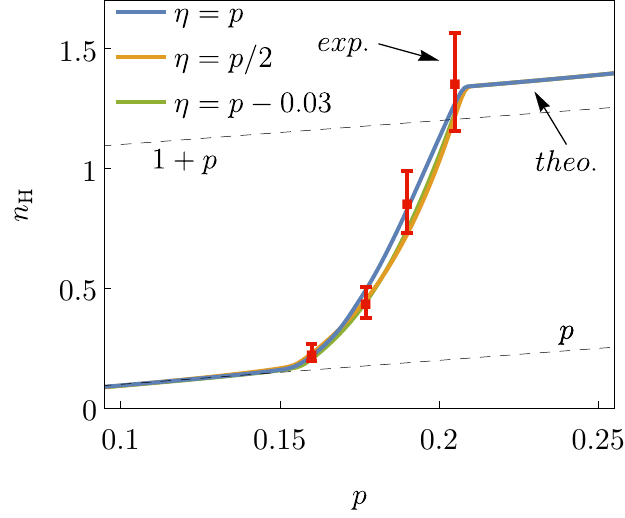} 
\caption{Fit of the Hall number as a function of doping to the experimental data from Badoux {\it et al.} \cite{badoux16} For the relaxation rate chosen as $\Gam/t = 0.05$, best agreement for $\eta = p$ is obtained for $p^* = 0.21$ and $D/t = 16.5$. Also shown are results obtained with the same parameters but $\eta(p) = p/2$ and $\eta(p) = p - 0.03$.}
\label{fig_fit}
\end{figure}

So far we have shown results for $p^* = 0.19$, the onset doping for the Hall number drop extracted from the experimental data by Badoux {\it et al.},\cite{badoux16} and $D/t = 12$ for the arbitrarily chosen prefactor of the doping dependent magnetic gap. The Hall number drop from values above $1+p$ for large doping to $p$ for small doping was reproduced by our results, if $\Gam/t$ is not too large.
To make closer contact to the experiment, we have fitted the parameters $p^*$ and $D$ to obtain quantitative agreement with the observed data points for YBCO. The fit, obtained for a fixed $\Gam/t = 0.05$, and shown in Fig.~\ref{fig_fit}, is optimal for $p^* = 0.21$ and $D/t = 16.5$.
Here we also compare to results obtained with the same values of $p^*$ and $D$, but with a different choice of the incommensurability $\eta(p)$, namely $\eta = p/2$ and $\eta = p - 0.03$. These alternative functions are closer to the incommensurabilities observed for YBCO. \cite{haug10} While the doping dependence of the Fermi-surface topologies depends on the choice of $\eta(p)$, one can see that the doping dependence of the Hall number is only weakly affected. 

The value of $D$ is unreasonably large. For a hopping amplitude $t \approx 0.3 \, \text{eV}$, the magnetic gap $\Delta(p) = D (p^* - p)$ would rise to a value $\Delta \approx 0.5 \, \text{eV}$ at $p = 0.1$. Large values for $D$ were also assumed in previous studies of the Hall effect in N\'eel and spiral antiferromagnetic states, to obtain a sufficiently steep decrease of the Hall number. \cite{storey16,eberlein16,verret17}
The required size of $D$ can be substantially reduced, if the bare hopping $t$ is replaced by a smaller effective hopping
\begin{equation}
 t_{\rm eff} = \frac{2p}{1+p} \, t \, ,
\end{equation}
where the Gutzwiller factor on the right-hand side captures phenomenologically the loss of metallicity in the doped Mott insulator. Such a factor is used in the YRZ-ansatz for the pseudogap phase. \cite{yang06}
Replacing $t$ by $t_{\rm eff}$ with $t = 0.3 \, \text{eV}$, a prefactor $D = 1.5 \, \text{eV}$ is sufficient to obtain the best fit for $n_H$, leading to $\Delta \approx 0.15 \, \text{eV}$ at $p=0.1$. This value is similar to the magnetic energy scale $J$ in cuprates.

All our results have been computed by evaluating the conductivity formulas with a Fermi function at zero temperature. We have checked that the temperature dependence from the Fermi function is negligible at the temperatures at which the recent transport experiments in cuprates \cite{badoux16,collignon17} have been carried out.

%%% Conclusion %%%%%%%%%%%%%%%%%%%%%%%%%%%%%%%%%%%%%%%%%%%%%%%%%%%%%%%%%%%

\section{Conclusion}

We have computed electrical dc conductivities in a two-dimensional metal with spiral magnetic order. Scattering processes were modeled by a momentum-independent relaxation rate $\Gam$. We have derived an expression for the longitudinal conductivity and a complete formula (including all interband contributions) for the Hall conductivity in the low-field limit $\om_c \tau \ll 1$.

For small $\Gam$, interband terms are suppressed by a factor of order $\Gam^2$ compared to the dominant intraband contributions. In the limit $\Gam \to 0$, the interband contributions are negligible and the intraband contributions simplify to the formulas derived by Voruganti {\it et al.} \cite{voruganti92} The latter have the same structure as the conductivities for non-interacting electrons in relaxation time approximation, with the bare electron dispersion $\eps_\bp$ replaced by the quasiparticle dispersions $E_\bp^\pm$ in the spiral state. We expect that this is true for any charge or spin density wave state in mean-field theory. This is also suggested by a recent general analysis of charge transport in interacting multiband systems with arbitrary band topology. \cite{nourafkan18} 

A numerical evaluation of the conductivities for band parameters as in YBCO and various choices of the relaxation rate $\Gam$ shows that interband contributions start playing a significant role only for $\Gam/t > 0.1$, where $t$ is the nearest-neighbor hopping amplitude. Relaxation rates observed in recent high-field transport experiments for cuprates are smaller, so that the application of the simple formulas derived by Voruganti {\it et al.} \cite{voruganti92} is justified.

The magnetic order induces a reduction of the longitudinal conductivity and of the Hall number.
The longitudinal conductivity in the spiral state exhibits a pronounced doping dependent nematicity in agreement with experimental observations in cuprates. \cite{ando02}
With a doping dependent magnetic gap of the form $\Delta(p) = D (p^* - p)$ for $p < p^*$, the Hall number drop below the critical doping $p^*$ observed in experiments \cite{badoux16,collignon17} can be well described. To fit the experimental data with a realistic (not too large) value of $D$, the reduction of the hopping amplitudes by correlation effects has to be taken into account.

Spiral magnetic order is thus consistent with transport experiments in cuprates, where superconductivity is suppressed by high magnetic fields. We finally note that fluctuating instead of static magnetic order should yield similar transport properties, as long as pronounced magnetic correlations are present.
% \\
%%%%%%%%%%%%%%%%%%%%%%%%%%%%%%%%%%%%%%%%%%%%%%%%%%%%%%%%%%%%%%%%%%%%%%%%%%
\begin{acknowledgments}
We are grateful to A.~Eberlein, R.~Nourafkan, J.~Schmalian, O.~Sushkov, L.~Taillefer, A.-M.~Tremblay, H.~Yamase, and R.~Zeyher for valuable discussions. 
\end{acknowledgments}

%%% Appendices %%%%%%%%%%%%%%%%%%%%%%%%%%%%%%%%%%%%%%%%%%%%%%%%%%%%%%%%%%%
\onecolumngrid
% \appendix

\begin{appendix}
% \begin{widetext}
%%%%%%%%%%%%%%%%%%%%%%%%%%%%%%%%%%%%%%%%%%%%%%%%%%%%%%%%%%%%%%%%%%%%%%%%%%%%%%%

\section{Analytic continuation} \label{AnalyticContinuation}

For the analytic continuation of the response functions we use the spectral representation of the imaginary frequency propagator
\begin{equation}
\label{SpectralG}
 \cG_{ip_0,\bp} =
 \int_{-\infty}^\infty d\eps \, \frac{\cA_\bp(\eps)}{ip_0 - \eps} \, ,
\end{equation}
where
\begin{equation}
 \cA_\bp(\eps) = \left( \begin{array}{cc}
 A^+_\bp(\eps) & 0 \\[0mm] 0 & A^-_\bp(\eps) \end{array} \right)
\end{equation}
is the matrix of spectral functions of the two quasiparticle bands. For simplicity of notation, we drop the momentum dependence in the following.
Real frequency quantities are conveniently formulated in terms of advanced and retarded Green functions,
\begin{equation}
 \cG_\eps^A =
 \int_{-\infty}^\infty d\eps' \, \frac{\cA(\eps')}{\eps - \eps' - i0^+} \, ,
\end{equation}
\begin{equation}
 \cG_\eps^R =
 \int_{-\infty}^\infty d\eps' \, \frac{\cA(\eps')}{\eps - \eps' + i0^+} \, .
\end{equation}
The functions to be continued analytically have the following structure
\begin{equation}
 I_{iq_0}^{m,n} = T \sum_{p_0} {\rm tr} \big(
 \cG_{ip_0+iq_0} \,\cM_1 \dots \cG_{ip_0+iq_0} \,\cM_m \cG_{ip_0} \,\cN_1 \dots \cG_{ip_0} \,\cN_n
 \big) \, ,
\end{equation}
where $\cM_1,\dots,\cM_m$ and $\cN_1,\dots,\cN_n$ are frequency-independent $2 \times 2$ matrices, and $q_0$ is a bosonic Matsubara frequency.
We insert the spectral representation \eqref{SpectralG} for each propagator and
perform the Matsubara frequency sum over the resulting product of energy denominators.  
Using $\text{Res}_{ip_0}[f(\eps)] = -T$, where $f(\eps)=(e^{\eps/T}+1)^{-1}$ is the Fermi function, we apply the residue theorem to replace the Matsubara frequency sum by a contour integral encircling the fermionic Matsubara frequencies counterclockwise. We then change the contour such that only the poles from the energy denominators are encircled. Applying the residue theorem again yields
\begin{eqnarray}
 && T \sum_{p_0} \frac{1}{ip_0 + iq_0 - \eps_1} \dots \frac{1}{ip_0 + iq_0 - \eps_m}
 \, \frac{1}{ip_0 - \eps'_1} \dots \frac{1}{ip_0 - \eps'_n}
 \nonumber \\
 && = f(\eps_1) \frac{1}{\eps_1 - \eps_2} \dots \frac{1}{\eps_1 - \eps_m} \,
 \frac{1}{-iq_0 + \eps_1 - \eps'_1} \dots \frac{1}{-iq_0 + \eps_1 - \eps'_n} + \dots
 \nonumber \\
 && + \, f(\eps_m) \frac{1}{\eps_m - \eps_1} \dots \frac{1}{\eps_m - \eps_{m-1}} \,
 \frac{1}{-iq_0 + \eps_m - \eps'_1} \dots \frac{1}{-iq_0 + \eps_m - \eps'_n}
 \nonumber \\
 && + \, f(\eps'_1) \frac{1}{iq_0 + \eps'_1 - \eps_1} \dots 
 \frac{1}{iq_0 + \eps'_1 - \eps_m} \,
 \frac{1}{\eps'_1 - \eps'_2} \dots \frac{1}{\eps'_1 - \eps'_n} + \dots
 \nonumber \\
 && + \, f(\eps'_n) \frac{1}{iq_0 + \eps'_n - \eps_1} \dots 
 \frac{1}{iq_0 + \eps'_n - \eps_m} \,
 \frac{1}{\eps'_n - \eps'_1} \dots \frac{1}{\eps'_n - \eps'_{n-1}} \, .
\end{eqnarray}
This expression can be easily continued to real frequencies, replacing $iq_0$ by $\om + i0^+$. Performing the integrals over $\eps_1,\dots,\eps_m$ and $\eps'_1,\dots,\eps'_n$ then yields
\begin{align} \label{Imn}
 I_\om^{m,n} =& \int d\eps f(\eps) \, {\rm tr} \bigg\{
 \Big[ \cA(\eps) \cM_1 \cP(\eps) \cM_2 \dots \cP(\eps) \cM_m + \dots
 + \, \cP(\eps) \cM_1 \dots \cP(\eps) \cM_{m-1} \cA(\eps) \cM_m \Big]
 \cG_{\eps-\om}^A\, \cN_1 \dots \cG_{\eps-\om}^A\, \cN_n \bigg\}
 \nonumber \\
 &+ \int d\eps f(\eps) \, {\rm tr} \bigg\{
 \cG_{\eps+\om}^R \,\cM_1 \dots \cG_{\eps+\om}^R \,\cM_m \Big[
  \cA(\eps) \cN_1 \cP(\eps) \cN_2 \dots \cP(\eps) \cN_n+ \dots
 + \, \cP(\eps) \cN_1 \dots \cP(\eps) \cN_{n-1} \cA(\eps) \cN_n \Big] \bigg\} \, ,
\end{align}
where $\cP(\eps)$ is the principal value integral
\begin{equation}
 \cP(\eps) = {\rm P.V.} \! \int d\eps' \, \frac{\cA(\eps')}{\eps - \eps'} =
 \frac{1}{2} \big( \cG_\eps^A + \cG_\eps^R \big) \, .
\end{equation}
%

%%%%%%%%%%%%%%%%%%%%%%%%%%%%%%%%%%%%%%%%%%%%%%%%%%%%%%%%%%%%%%%%%%%%%%%%%%%%

\section{Evaluation of ordinary conductivity} \label{OrdinaryConductivity}

We first present the derivation leading from Eq.~(\ref{K1}) to Eq.~(\ref{K2}). 
Splitting the vertices in purely diagonal and off-diagonal contributions, one obtains
\begin{equation} \label{K1a}
 K_{iq_0}^{\alf\beta} = e^2 \trp \big(
 \cG_{\bp,ip_0+iq_0} \cE_\bp^\alf \cG_{\bp,ip_0} \cE_\bp^\beta +
 \cG_{\bp,ip_0+iq_0} \cF_\bp^\alf \cG_{\bp,ip_0} \cF_\bp^\beta +
 \cG_{\bp,ip_0} \cE_\bp^{\alf\beta} - \cG_{\bp,ip_0} \cC_\bp^{\alf\beta}
 \big) \, ,
\end{equation}
since only terms with an even number of off-diagonal matrices contribute to the trace.
Here and in the following we use the shorthand notation
$\trp = T L^{-1} \sum_p {\rm tr}(\dots)$.
Using a partial integration, one obtains the identity
\begin{equation}
 \trp \big( \cG_{\bp,ip_0} \cE_\bp^{\alf\beta} \big) =
 \trp \big( \cG_{\bp,ip_0} \partial_{p_\alf} \cE_\bp^\beta \big) =
 - \trp \big[ (\partial_{p_\alf} \cG_{\bp,ip_0}) \cE_\bp^\beta \big] =
 - \trp \big( \cG_{\bp,ip_0} \cE_\bp^\alf \cG_{\bp,ip_0} \cE_\bp^\beta \big)
 \, .
\end{equation}
A few purely algebraic steps yield the relation
\begin{equation}
 \trp \big( \cG_{\bp,ip_0} \cC_\bp^{\alf\beta} \big) =
 \trp \big( \cG_{\bp,ip_0} \cF_\bp^\alf \cG_{\bp,ip_0} \cF_\bp^\beta \big) \, .
\end{equation}
Hence, the last two (diamagnetic) contributions in Eq.~(\ref{K1a}) cancel the first two (paramagnetic) contributions for $q_0 = 0$, and we obtain Eq.~(\ref{K2}).

We now perform the Matsubara sum and the analytic continuation to real frequencies. The frequency dependence in Eq.~(\ref{K2}) has the form
\begin{equation}
 K_{iq_0} = T \sum_{p_0} {\rm tr}
 \big[ (\cG_{ip_0+iq_0} - \cG_{ip_0}) \cM_1 \cG_{ip_0} \cM_2 \big] \, ,
\end{equation}
where $\cM_1$ and $\cM_2$ are frequency-independent matrices. We have dropped the momentum dependence, since it does not interfere with the following steps.
Applying the general formula Eq.~(\ref{Imn}) in Appendix \ref{AnalyticContinuation} for the cases $m=n=1$ and $m=0$, $n=2$ yields
\begin{equation}
 K_\om = \int d\eps f(\eps) \, {\rm tr} \big[
 \cM_1 \cA(\eps) \cM_2 (\cG_{\eps-\om}^A - \cG_\eps^A) +
 \cM_1 (\cG_{\eps+\om}^R - \cG_\eps^R) \cM_2 \cA(\eps) \big] \, .
\end{equation}
Using the cyclic property of the trace, and the fact that all involved matrices are invariant under the exchange of row and column indices, this can also be written as
\begin{equation}
 K_\om = \int d\eps f(\eps) {\rm tr} \, \big[ \cM_1 \cA(\eps) \cM_2
 (\cG_{\eps-\om}^A - \cG_\eps^A + \cG_{\eps+\om}^R - \cG_\eps^R) \big] \, .
\end{equation}

For the dc conductivity, we need to compute the ratio $K_\om/i\om$ in the limit $\om \to 0$. Using $(\cG_{\eps-\om}^A - \cG_\eps^A)/\om \to - \partial_\eps \cG_\eps^A$ and $(\cG_{\eps+\om}^R - \cG_\eps^R)/\om \to \partial_\eps \cG_\eps^R$ for $\om \to 0$, and the relation $\cG_\eps^R - \cG_\eps^A = -2\pi i \cA(\eps)$, one obtains
\begin{equation}
 \lim_{\om \to 0} \frac{K_\om}{i\om} =
 - 2\pi \int d\eps f(\eps) {\rm tr}
 \big[ \cM_1 \cA(\eps) \cM_2 \partial_\eps \cA(\eps) \big] \, .
\end{equation}
Using the relation ${\rm tr} \big[ \cM_1 \cA(\eps) \cM_2 \partial_\eps \cA(\eps) \big] =
{\rm tr} \big[ \cM_1 \partial_\eps \cA(\eps) \cM_2 \cA(\eps) \big]$ and a partial integration, one can shift the frequency derivative on the Fermi function,
\begin{equation} \label{Iom1}
 \lim_{\om \to 0} \frac{K_\om}{i\om} =
 \pi \int d\eps f'(\eps) {\rm tr}
 \big[ \cM_1 \cA(\eps) \cM_2 \cA(\eps) \big] \, .
\end{equation}
Applying this result to Eq.~(\ref{K2}) yields the conductivity in the form
\begin{equation}
 \sg^{\alf\beta} = - \lim_{\om \to 0} \frac{K_\om^{\alf\beta}}{i\om} =
 - e^2 \frac{\pi}{L} \sum_\bp \int d\eps f'(\eps) {\rm tr} \big[
 \cE_\bp^\alf \cA_\bp(\eps) \cE_\bp^\beta \cA_\bp(\eps) +
 \cF_\bp^\alf \cA_\bp(\eps) \cF_\bp^\beta \cA_\bp(\eps) \big] \, .
\end{equation}
Performing the matrix products and the trace one obtains the formula (\ref{sg}) presented in the main text.

%%%%%%%%%%%%%%%%%%%%%%%%%%%%%%%%%%%%%%%%%%%%%%%%%%%%%%%%%%%%%%%%%%%%%%%%%%%%%%%%%%

\section{Evaluation of Hall conductivity} \label{HallConductivity}

%%%%%%%%%%%%%%%%%%%%%%%%%%%%%%%%%%%%%%%%%%%%%%%%%%%%%%%%%%%%%%%%%%%%%%%%%%%%%%%%%%

\subsection{Vanishing $K^{\alf\beta\gam}_{\bq,iq_0}$ for $\bq=0$}

The term $K^{\alf\beta\gam}_{\bq,iq_0}$ in Eq.~\eqref{K_H} should vanish for $\bq=0$ as a 
consequence of gauge invariance. We show that $K^{\alf\beta\gam}_{\bq=0,iq_0}$ indeed vanishes by explicit calculation.
$K^{\alf\beta\gam}_{\bq,iq_0}$ in Eq.~\eqref{K_H} for $\bq=0$ reads
\begin{eqnarray} \label{K_H0}
 K_{\bq=0,iq_0}^{\alf\beta\gam} &=& e^3 \, \trp\left[
 G_{\bp,ip_0} \lam_{\bp\bp}^{\alf\beta\gam}
 + G_{\bp,ip_0} \lam_{\bp\bp}^\gamma
     G_{\bp,ip_0} \lam_{\bp\bp}^{\alf\beta} \label{K_H01}
 \right.\\
  &+& \left. G_{\bp,ip_0+iq_0} \lam_{\bp\bp}^\beta
     G_{\bp,ip_0} \lam_{\bp\bp}^{\alf\gam}
 + G_{\bp,ip_0+iq_0} \lam_{\bp\bp}^\alf
     G_{\bp,ip_0} \lam_{\bp\bp}^{\beta\gam} \label{K_H02}
 \right. \\
  &+& \left.G_{\bp,ip_0+iq_0} \lam_{\bp\bp}^\alf
     G_{\bp,ip_0} \lam_{\bp\bp}^\gam
     G_{\bp,ip_0} \lam_{\bp\bp}^\beta 
 + G_{\bp,ip_0-iq_0} \lam_{\bp\bp}^\alf
     G_{\bp,ip_0} \lam_{\bp\bp}^\gam
     G_{\bp,ip_0} \lam_{\bp\bp}^\beta \right] \, . \label{K_H03}
\end{eqnarray}
Here and in the following we use the shorthand notation $\trp=TL^{-1}\sum_p\text{tr}(\dots)$. We use the definition of the vertex $\lam^{\alf\beta\gam}_{\bp\bp}$ in Eq.~\eqref{lampp'}, perform a partial integration in $p^\gam$, and use that $\partial_\gam G_{\bp,ip_0}=G_{\bp,ip_0}\lam^\gam_{\bp\bp}G_{\bp,ip_0}$, which follows directly by 
definition in Eq.~\eqref{G_p}. We get
\begin{align}
 \trp\left[G_{\bp,ip_0}\lam^{\alf\beta\gam}_{\bp\bp}\right]&= \trp\left[G_{\bp,ip_0}\big(\partial_\gam\lam^{\alf\beta}_{\bp\bp}\big)\right] \nonumber\\
 &= -\trp\left[\big(\partial_\gam G_{\bp,ip_0}\big)\lam^{\alf\beta}_{\bp\bp}\right]=- \trp\left[G_{\bp,ip_0}\lam^\gam_{\bp\bp}G_{\bp,ip_0}\lam^{\alf\beta}_{\bp\bp}\right] \,  .
\end{align}
Thus, line \eqref{K_H01} vanishes. Performing the same steps on $G_{\bp,ip_0+iq_0} \lam_{\bp\bp}^\beta G_{\bp,ip_0} \lam_{\bp\bp}^{\alf\gam}$
in line \eqref{K_H02} leads to the remaining three contributions in \eqref{K_H02} and \eqref{K_H03} with opposite sign, after 
reversing the matrix order under the trace\cite{footnote2} and shifting the Matsubara summation. Thus, $K_{\bq=0,iq_0}^{\alf\beta\gam}$
vanishes.

%%%%%%%%%%%%%%%%%%%%%%%%%%%%%%%%%%%%%%%%%%%%%%%%%%%%%%%%%%%%%%%%%%%%%%%%%%%%%%%%%%%%

\subsection{Derivation leading from Eq.~\eqref{K_H} to Eq.~\eqref{Kabcd}}

We now present the derivation leading from Eq.~\eqref{K_H} to Eq.~\eqref{Kabcd}.
As shown in the main text,
$K_{\bq,iq_0}^{\alf\beta\gam\delta} = \partial_{q^\delta} K^{\alf\beta\gam}_{\bq,iq_0}$
with $K^{\alf\beta\gam}_{\bq,iq_0}$ from Eq.~\eqref{K_H} reduces to three contributions in a uniform magnetic field ($\bq\rightarrow 0$):
\begin{eqnarray} \label{K_Hbare}
 K_{iq_0}^{\alf\beta\gam\delta} &=& e^3 \frac{\partial}{\partial q^\delta}\trp\left.\left[
 G_{\bp^+,ip_0+iq_0} \lam_{\bp^+\bp^-}^\alf
     G_{\bp^-,ip_0} \lam_{\bp^-\bp^+}^{\beta\gam}
 \nonumber \right.\right.\\
 &+& \left.\left.G_{\bp^+,ip_0+iq_0} \lam_{\bp^+\bp^-}^\alf
     G_{\bp^-,ip_0} \lam_{\bp^-\bp^+}^\gam
     G_{\bp^+,ip_0} \lam_{\bp^+\bp^+}^\beta 
 \nonumber \right.\right.\\
 &+& \left.\left.G_{\bp^-,ip_0-iq_0} \lam_{\bp^-\bp^+}^\alf
     G_{\bp^+,ip_0} \lam_{\bp^+\bp^-}^\gam
     G_{\bp^-,ip_0} \lam_{\bp^-\bp^-}^\beta \right]\right|_{\bq=0} \, ,
\end{eqnarray}
with the shorthand notation $\bp^\pm=\bp\pm\frac{1}{2}\bq$. 

It turns out to be more convenient to perform the derivative in the nonrotated basis with the Green function $G_p$ in Eq.~\eqref{G_p} and the vertices $\lam^{\alf_1\dots\alf_n}_{\bp\bp'}$ in Eq.~\eqref{lampp'}. \cite{nourafkan18}
Using $\partial_\delta \left.G_{\bp^\pm,p_0}\right|_{\bq=0}=\mp\frac{1}{2}G_{\bp,p_0}\big(\partial_\delta \,G^{-1}_{\bp,p_0}\big)G_{\bp,p_0}$,
which follows from differentiating $G_p G^{-1}_p=1$, one gets
\begin{align}
 \partial_\delta \left.G_{\bp^\pm,p_0}\right|_{\bq=0}=\pm\frac{1}{2}G_{\bp,p_0}\lambda^\delta_{\bp\bp}G_{\bp,p_0}\, .
\end{align}
Using $\partial_\delta\left.\lam^{\alf_1\dots\alf_n}_{\bp+a\bq,\bp+b\bq}\right|_{\bq=0}=\frac{a+b}{2}\lam^{\alf_1\dots\alf_n\delta}_{\bp\bp}$, 
one immediately notices that only 
\begin{align}
 \partial_\delta\left.\lam_{\bp^\pm\bp^\pm}^\beta\right|_{\bq=0}=\pm\frac{1}{2}\lam^{\beta\delta}_{\bp\bp} 
\end{align}
has a nonzero derivative.

We combine the contributions from the derivative of the two Green functions in the first line in Eq.~\eqref{K_Hbare} with the derivative of the vertices in the second and third lines in Eq.~\eqref{K_Hbare}. After shifting the Matsubara summation by $iq_0$ and reversing the matrix order under the trace\cite{footnote2} of the latter one, we obtain
\begin{align} \label{KH1}
 (\Kabgd)^{(1)}=\frac{e^3}{2}\trp\left[G_{\bp,ip_0+iq_0}\lam^\alf_{\bp\bp}G_{\bp,ip_0}\lam^\gam_{\bp\bp}G_{\bp,ip_0}\lam^{\beta\delta}_{\bp\bp}-(iq_0\rightarrow -iq_0)-(\gamma\leftrightarrow \delta)\right] \, .
\end{align}
Note that the expression is antisymmetric when replacing $iq_0\rightarrow -iq_0$ and exchanging $\gamma\leftrightarrow\delta$,
which are both required by gauge invariance. 

We are left with the six derivatives of the Green functions in the second and third lines in Eq.~\eqref{K_Hbare}. After shifting the
Matsubara summation by $iq_0$ and reversing the matrix order under the trace we get
\begin{align} \label{KH2}
 (\Kabgd)^{(2)}=\frac{e^3}{2}\trp\left[G_{\bp,ip_0+iq_0}\lam^\alf_{\bp\bp}G_{\bp,ip_0}\lam^\gam_{\bp\bp}G_{\bp,ip_0}\lam^\delta_{\bp\bp}G_{\bp,ip_0}\lam^\beta_{\bp\bp}-(iq_0\rightarrow -q_0)-(\gamma\leftrightarrow\delta)\right]
\end{align}
and
\begin{align} \label{KH3}
 (\Kabgd)^{(3)}=\frac{e^3}{4}\trp\left[G_{\bp,ip_0+iq_0}\lam^\delta_{\bp\bp}G_{\bp,ip_0+iq_0}\lam^\alf_{\bp\bp}G_{\bp,ip_0}\lam^\gamma_{\bp\bp}G_{\bp,ip_0}\lam^\beta_{\bp\bp}-(iq_0\rightarrow -q_0)-(\gamma\leftrightarrow\delta)\right]\, .
\end{align}
Again, the expressions are antisymmetric when replacing $iq_0 \rightarrow -iq_0$ and exchanging $\gamma \leftrightarrow \delta$, as required by gauge invariance. 

%%%%%%%%%%%%%%%%%%%%%%%%%%%%%%%%%%%%%%%%%%%%%%%%%%%%%%%%%%%%%%%%%%%%%%%%%%%%%%%%

\subsubsection{Antisymmetry in $\alf\leftrightarrow\beta$}

We expect the result to be antisymmetric under the exchange of the indices $\alf\leftrightarrow\beta$. 
Let us start with $(\Kabgd)^{(2)}$ in Eq.~\eqref{KH2}. We split it in two equal parts and reverse the matrix order under the trace of the second term. 
After changing sign by $(\gamma\leftrightarrow\delta)$ we obtain the antisymmetric counterpart of the first term. Thus, we get
\begin{align} \label{KHrec}
 (\Kabgd)^{(\text{rec})} = \frac{e^3}{4}\widetrp\left[\cG_{\bp,ip_0+iq_0}\tlam^\alf_{\bp\bp}\cG_{\bp,ip_0}\tlam^\gam_{\bp\bp}\cG_{\bp,ip_0}\tlam^\delta_{\bp\bp}\cG_{\bp,ip_0}\tlam^\beta_{\bp\bp}\right] \, ,
\end{align}
where we introduced the short notation 
\begin{align} \label{widetildetrp}
 \widetrp[\cdots]=\trp\left[\dots -(iq_0\rightarrow -iq_0)-(\gam\leftrightarrow \delta)-(\alf\leftrightarrow\beta)\right] \, .
\end{align}
We changed back to the quasiparticle basis. The upper index ``rec'' indicates that it corresponds to a rectangular diagram in analogy with the terminology of Ref.~\onlinecite{nourafkan18}.
Note that we do not identify the other contribution with four vertices $(\Kabgd)^{(3)}$ as a rectangular diagram, in contrast to Ref.~\onlinecite{nourafkan18}, for the reason explained below.

Let us continue with $(\Kabgd)^{(3)}$ in Eq.~\eqref{KH3}. We can reintroduce two derivatives by using the relation $\partial_\delta G_p=G_p\lambda^\delta_{\bp\bp}G_p$:
\begin{align}
  \frac{e^3}{4}\trp\left[\left(\partial_\delta G_{\bp,ip_0+iq_0}\right)\lam^\alf_{\bp\bp}\left(\partial_\gamma G_{\bp,ip_0}\right)\lam^\beta_{\bp\bp}-(iq_0\rightarrow -q_0)-(\gamma\leftrightarrow\delta)\right]\, .
\end{align}
We perform partial integration in $p^\delta$. The term including $\partial_\delta\partial_\gamma G_{\bp,ip_0}$ drops out due to its counterpart in $(\gamma\leftrightarrow\delta)$.
After reversing the matrix order under the trace we obtain
\begin{align}
  (\Kabgd)^{(3)}=&-\frac{e^3}{4}\trp\left[G_{\bp,ip_0+iq_0}\lam^\beta_{\bp\bp}G_{\bp,ip_0}\lam^\gamma_{\bp\bp} G_{\bp,ip_0}\lam^{\alf\delta}_{\bp\bp}-(iq_0\rightarrow -q_0)-(\gamma\leftrightarrow\delta)\right]\\
  &-\frac{e^3}{4}\trp\left[G_{\bp,ip_0+iq_0}\lam^\alf_{\bp\bp}G_{\bp,ip_0}\lam^\gamma_{\bp\bp}G_{\bp,ip_0}\lam^{\beta\delta}_{\bp\bp}-(iq_0\rightarrow -q_0)-(\gamma\leftrightarrow\delta)\right]\, .
\end{align}
Combining this with $(\Kabgd)^{(1)}$ in Eq.~\eqref{KH1} we end up with
\begin{align} \label{KHtri}
 (\Kabgd)^{(\text{tri})}=\frac{e^3}{4}\widetrp\left[\cG_{\bp,ip_0+iq_0}\tlam^\alf_{\bp\bp}\cG_{\bp,ip_0}\tlam^\gam_{\bp\bp}\cG_{\bp,ip_0}\tlam^{\beta\delta}_{\bp\bp}\right] \, ,
\end{align}
where we changed back to the quasiparticle basis. The trace was defined in Eq.~\eqref{widetildetrp}. The upper index ``tri'' indicates that it is a triangular diagram 
by analogy with the terminology of Ref.~\onlinecite{nourafkan18}.

%%%%%%%%%%%%%%%%%%%%%%%%%%%%%%%%%%%%%%%%%%%%%%%%%%%%%%%%%%%%%%%%%%%%%%%%%%%%%%%%

\subsubsection{Decomposition of $(\Kabgd)^{(\text{tri})}$}

In the following simplification, we will use the explicit matrix form of the various components. The Green functions $\cG$ are already diagonal.
In the main text we introduced the decomposition $\tlam^\alf_{\bp\bp}=\cE^\alf_\bp+\cF^\alf_\bp$ and
$\tlam^{\alf\beta}_{\bp\bp}=\cE^{\alf\beta}_\bp-\cC^{\alf\beta}_\bp+\cH^{\alf\beta}_\bp$, where $\cE^\alf_\bp,\,\cE^{\alf\beta}_\bp$ and $\cC^{\alf\beta}_\bp$ are diagonal matrices, whereas $\cF^\alf_\bp$ and $\cH^{\alf\beta}_\bp$ are off-diagonal matrices.

Using that only an even number of off-diagonal matrices contribute under the trace, the contribution $(\Kabgd)^{(\text{tri})}$ in Eq.~\eqref{KHtri} decomposes in six contributions, which we label as
\begin{align}
 (\Kabgd)^{(\text{tri},\text{I})}&=+\frac{e^3}{4}\widetrp\left[
 \cG_{\bp,ip_0+iq_0}\cE^{\alf}_{\bp}\cG_{\bp,ip_0}\cE^\gam_\bp \cG_{\bp,ip_0}\cE^{\beta\delta}_\bp\right]\label{KH1i}\, ,\\
 (\Kabgd)^{(\text{tri},\text{II})}&=-\frac{e^3}{4}\widetrp\left[
 \cG_{\bp,ip_0+iq_0}\cE^{\alf}_{\bp}\cG_{\bp,ip_0}\cE^\gam_\bp \cG_{\bp,ip_0}\cC^{\beta\delta}_\bp\right]\label{KH1ii}\, ,\\
 (\Kabgd)^{(\text{tri},\text{III})}&=+\frac{e^3}{4}\widetrp\left[
 \cG_{\bp,ip_0+iq_0}\cF^{\alf}_{\bp}\cG_{\bp,ip_0}\cE^\gam_\bp \cG_{\bp,ip_0}\cH^{\beta\delta}_\bp\right]\label{KH1iii}\, ,\\
 (\Kabgd)^{(\text{tri},\text{IV})}&=+\frac{e^3}{4}\widetrp\left[
 \cG_{\bp,ip_0+iq_0}\cE^\alf_\bp\cG_{\bp,ip_0}\cF^\gam_\bp\cG_{\bp,ip_0}\cH^{\beta\delta}_\bp\right]\label{KH1iv}\, , \\
 (\Kabgd)^{(\text{tri},\text{V})}&=+\frac{e^3}{4}\widetrp\left[
 \cG_{\bp,ip_0+iq_0}\cF^\alf_\bp\cG_{\bp,ip_0}\cF^\gam_\bp\cG_{\bp,ip_0}\cE^{\beta\delta}_\bp\right]\label{KH1v} \, , \\
  (\Kabgd)^{(\text{tri},\text{VI})}&=-\frac{e^3}{4}\widetrp\left[
 \cG_{\bp,ip_0+iq_0}\cF^\alf_\bp\cG_{\bp,ip_0}\cF^\gam_\bp\cG_{\bp,ip_0}\cC^{\beta\delta}_\bp\right]\label{KH1vi} \, .
\end{align}
The last one cancels by antisymmetry in $\alf\leftrightarrow\beta$,
\begin{align}
 (\Kabgd)^{(\text{tri},\text{VI})}=0 \, ,
\end{align}
when using the relation $\cC^{\beta\delta}_\bp=2\,\sgc\cF^\beta_\bp\cF^\delta_\bp$,
which follows from the definition Eq.~\eqref{Cp} and $\sgc=1/(E^+_\bp-E^-_\bp)\,\sigma^z$.

%%%%%%%%%%%%%%%%%%%%%%%%%%%%%%%%%%%%%%%%%%%%%%%%%%%%%%%%%%%%%%%%%%%%%%%%%%%%%%%%

\subsubsection{Decomposition of $(\Kabgd)^{(\text{rec})}$}
Using that only an even number of off-diagonal matrices contribute under the trace, the contribution
$(\Kabgd)^{(\text{rec})}$ in Eq.~\eqref{KHrec} decomposes in eight contributions, which we label as
\begin{align}
 (\Kabgd)^{(\text{\text{rec},I})}&=\frac{e^3}{4}\widetrp\left[\cG_{\bp,ip_0+iq_0}\cE^\alf_\bp\cG_{\bp,ip_0}\cE^\gam_\bp\cG_{\bp,ip_0}\cE^\delta_\bp\cG_{\bp,ip_0}\cE^\beta_\bp\right] \label{KH2i} \, ,\\
 (\Kabgd)^{(\text{\text{rec},II})}&=\frac{e^3}{4}\widetrp\left[\cG_{\bp,ip_0+iq_0}\cE^\alf_\bp\cG_{\bp,ip_0}\cE^\gam_\bp\cG_{\bp,ip_0}\cF^\delta_\bp\cG_{\bp,ip_0}\cF^\beta_\bp\right] \label{KH2ii} \, ,\\
 (\Kabgd)^{(\text{\text{rec},III})}&=\frac{e^3}{4}\widetrp\left[\cG_{\bp,ip_0+iq_0}\cE^\alf_\bp\cG_{\bp,ip_0}\cF^\gam_\bp\cG_{\bp,ip_0}\cE^\delta_\bp\cG_{\bp,ip_0}\cF^\beta_\bp\right] \label{KH2iii} \, ,\\
 (\Kabgd)^{(\text{\text{rec},IV})}&=\frac{e^3}{4}\widetrp\left[\cG_{\bp,ip_0+iq_0}\cE^\alf_\bp\cG_{\bp,ip_0}\cF^\gam_\bp\cG_{\bp,ip_0}\cF^\delta_\bp\cG_{\bp,ip_0}\cE^\beta_\bp\right] \label{KH2iv} \, ,\\
 (\Kabgd)^{(\text{\text{rec},V})}&=\frac{e^3}{4}\widetrp\left[\cG_{\bp,ip_0+iq_0}\cF^\alf_\bp\cG_{\bp,ip_0}\cE^\gam_\bp\cG_{\bp,ip_0}\cE^\delta_\bp\cG_{\bp,ip_0}\cF^\beta_\bp\right] \label{KH2v} \, ,\\
 (\Kabgd)^{(\text{\text{rec},VI})}&=\frac{e^3}{4}\widetrp\left[\cG_{\bp,ip_0+iq_0}\cF^\alf_\bp\cG_{\bp,ip_0}\cE^\gam_\bp\cG_{\bp,ip_0}\cF^\delta_\bp\cG_{\bp,ip_0}\cE^\beta_\bp\right] \label{KH2vi} \, ,\\
 (\Kabgd)^{(\text{\text{rec},VII})}&=\frac{e^3}{4}\widetrp\left[\cG_{\bp,ip_0+iq_0}\cF^\alf_\bp\cG_{\bp,ip_0}\cF^\gam_\bp\cG_{\bp,ip_0}\cE^\delta_\bp\cG_{\bp,ip_0}\cE^\beta_\bp\right] \label{KH2vii} \, , \\
 (\Kabgd)^{(\text{\text{rec},VIII})}&=\frac{e^3}{4}\widetrp\left[\cG_{\bp,ip_0+iq_0}\cF^\alf_\bp\cG_{\bp,ip_0}\cF^\gam_\bp\cG_{\bp,ip_0}\cF^\delta_\bp\cG_{\bp,ip_0}\cF^\beta_\bp\right] \label{KH2viii} \, .
\end{align}
The four contributions
\begin{align}
 (\Kabgd)^{(\text{\text{rec},I})} =
 (\Kabgd)^{(\text{\text{rec},IV})} =
 (\Kabgd)^{(\text{\text{rec},V})} =
 (\Kabgd)^{(\text{\text{rec},VIII})} = 0 
\end{align}
vanish due to their antisymmetric counterpart in $(\gamma\leftrightarrow\delta)$,
which follows from $\cE^\gam_\bp\cG_{\bp,ip_0}\cE^\delta_\bp=\cE^\delta_\bp\cG_{\bp,ip_0}\cE^\gam_\bp$ and $\cF^\gam_\bp\cG_{\bp,ip_0}\cF^\delta_\bp=\cF^\delta_\bp\cG_{\bp,ip_0}\cF^\gam_\bp$.
Furthermore, we have $(\Kabgd)^{(\text{\text{rec},II})}=(\Kabgd)^{(\text{\text{rec},VII})}$
and $(\Kabgd)^{(\text{\text{rec},III})}=(\Kabgd)^{(\text{\text{rec},VI})}$. 
In order to see this, reverse the matrix order under the trace and change both $\alf\leftrightarrow\beta$ and $\gamma\leftrightarrow\delta$.
We continue by applying the identity
\begin{align}
 \label{RedGreenFunc}
 \cG_{\bp,ip_0}\cF^\delta_\bp\cG_{\bp,ip_0}=\cF^\delta_\bp\cG_{\bp,ip_0}\sgc+\sgc\cG_{\bp,ip_0}\cF^\delta_\bp \, ,
\end{align}
which can be verified by purely algebraic steps with $\sgc=1/(E^+_\bp-E^-_\bp)\,\sigma^z$, to the remaining two contributions. We get
\begin{align}
 &+\frac{e^3}{2}\widetrp\left[\cG_{\bp,ip_0+iq_0}\cE^\alf_\bp\cG_{\bp,ip_0}\cE^\gam_\bp\cF^\delta_\bp\cG_{\bp,ip_0}\sgc\cF^\beta_\bp\right] \label{rec1}\\
 &+\frac{e^3}{2}\widetrp\left[\cG_{\bp,ip_0+iq_0}\cE^\alf_\bp\cG_{\bp,ip_0}\cE^\gam_\bp\sgc\cG_{\bp,ip_0}\cF^\delta_\bp\cF^\beta_\bp\right] \label{rec2}\\
 &+\frac{e^3}{2}\widetrp\left[\cG_{\bp,ip_0+iq_0}\cF^\alf_\bp\cG_{\bp,ip_0}\cE^\gam_\bp\cF^\delta_\bp\cG_{\bp,ip_0}\sgc\cE^\beta_\bp\right] \label{rec3}\\
 &+\frac{e^3}{2}\widetrp\left[\cG_{\bp,ip_0+iq_0}\cF^\alf_\bp\cG_{\bp,ip_0}\cE^\gam_\bp\sgc\cG_{\bp,ip_0}\cF^\delta_\bp\cE^\beta_\bp\right] \label{rec4}\, .
\end{align}
Let us first combine the two lines \eqref{rec1} and \eqref{rec3}. 
From the algebraic relation $G^+_p G^-_p =
\left(G^+_p - G^-_p \right)/\left(E^+_\bp-E^-_\bp\right)$ with
$G_p^\pm = [ip_0 + i\Gam\sgn(p_0) - E_\bp^\pm]^{-1}$
it immediately follows that the two types of Green-function products are equal under the Matsubara summation: 
\begin{align}
 &T\sum_{p_0}G^+_{\bp,ip_0}G^-_{\bp,ip_0}(G^+_{\bp,ip_0+iq_0}-G^+_{\bp,ip_0-iq_0})=T\sum_{p_0}G^+_{\bp,ip_0}G^-_{\bp,ip_0}(G^-_{\bp,ip_0+iq_0}-G^-_{\bp,ip_0-iq_0}) \, .
\end{align}
Thus, summing up the two lines \eqref{rec1} and \eqref{rec3} and performing the matrix trace explicitly leads to
\begin{align} \label{rec1+rec3}
 &\eqref{rec1}+\eqref{rec3}
 =\frac{e^3}{2} \, L^{-1}\sum_{\bp} \left[\frac{F^\alf_\bp F^\delta_\bp}{E^+_\bp-E^-_\bp}(E^{+,\gam}_\bp+E^{-,\gam}_{\bp}) (E^{+,\beta}_{\bp}-E^{-,\beta}_\bp) -(\alf\leftrightarrow\beta)-(\delta\leftrightarrow\gam)\right] \nonumber
 \\&\hspace{5cm}\times T \sum_{p_0} G^+_{\bp,ip_0} G^-_{\bp,ip_0} (G^+_{\bp,ip_0+iq_0}-G^+_{\bp,ip_0-iq_0})  \, .
\end{align}
From Eq.~\eqref{Ep} one obtains the derivative of the dispersion
$E^{\pm,\sg}_\bp=g^\sg_\bp\pm 2h_\bp h^\sg_\bp/(E^+_\bp-E^-_\bp)$.
Thus, with definition Eq.~\eqref{Fp} we have $E^{+,\sg}_\bp-E^{-,\sg}_\bp=\frac{2h_\bp}{\Delta}F^\sg_\bp$.
This immediately leads to 
\begin{align} \label{changingindices}
  F^\delta_\bp (E^{+,\beta}_{\bp}-E^{-,\beta}_\bp)=F^\beta_\bp (E^{+,\delta}_{\bp}-E^{-,\delta}_\bp) \,.
\end{align}
Then the bracket $\left[\cdots\right]$ in \eqref{rec1+rec3} vanishes by antisymmetry in $\alf\leftrightarrow\beta$.

We continue with \eqref{rec2}. We commute the two diagonal matrices $\sgc$ and $\cG_{\bp,ip_0}$. The last three 
matrices can then be combined by using the relation 
$\sgc\cF^\delta_\bp\cF^\beta_\bp=\frac{1}{2}\cC^{\beta\delta}_\bp$,
which follows from the definition Eq.~\eqref{Cp}. Thus, we get
\begin{align}
 \eqref{rec2}=\frac{e^3}{4}\widetrp\left[\cG_{\bp,ip_0+iq_0}\cE^\alf_\bp\cG_{\bp,ip_0}\cE^\gam_\bp\cG_{\bp,ip_0}\cC^{\beta\delta}_\bp\right]=-(\Kabgd)^{(\text{tri},\text{II})} \, ,
\end{align}
canceling \eqref{KH1ii}.

We close with \eqref{rec4}. We split it in two equal parts. In the first part, we reintroduce a derivative with respect to
$p^\gamma$ of a Green function after commuting the diagonal matrices $\sgc$ and $\cG_{\bp,ip_0}$:
\begin{align} \label{rec4part1}
 \frac{e^3}{4}\widetrp\left[\cG_{\bp,ip_0+iq_0}\cF^\alf_\bp\big(\partial_\gamma\cG_{\bp,ip_0}\big)\sgc\cF^\delta_\bp\cE^\beta_\bp\right] \, .
\end{align}
In the second part, we first shift the Matsubara summation $ip_0\rightarrow -ip_0$ and change the overall sign with its corresponding 
contribution in $(ip_0\rightarrow -ip_0)$. After reversing the matrix order under the trace, commuting $\sgc$ with $\cG_{\bp,ip_0+iq_0}$,
and reintroducing a derivative with respect to $p^\gamma$ of a Green function, we get
\begin{align} \label{rec4part2}
 -\frac{e^3}{4}\widetrp\left[\big(\partial_\gamma\cG_{\bp,ip_0+iq_0}\big)\cF^\alf_\bp\cG_{\bp,ip_0}\cE^\beta_\bp\cF^\delta_\bp\sgc\right] \, .
\end{align}
We use the identity $\cE^\beta_\bp\cF^\delta_\bp\sgc = \sgc\cF^\beta_\bp\cE^\delta_\bp+\cE^\delta_\bp\cF^\beta_\bp\sgc-\sgc\cF^\delta_\bp\cE^\beta_\bp$,
which immediately follows from \eqref{changingindices}. Only the last contribution is nonzero. The first two terms vanish by antisymmetry in $\alf\leftrightarrow\beta$. We sum up \eqref{rec4part1} and \eqref{rec4part2} and perform a partial integration in $p^\gamma$ in \eqref{rec4part1}. 
The term with a derivative acting on the Green function cancels \eqref{rec4part2}, and we obtain four contributions:
\begin{align} 
 \eqref{rec4part1}+\eqref{rec4part2} = &-\frac{e^3}{4}\widetrp\left[\cG_{\bp,ip_0+iq_0}\big(\partial_\gamma\cF^\alf_\bp\big)\cG_{\bp,ip_0}\sgc\cF^\delta_\bp\cE^\beta_\bp\right] \label{D51+D52part1} \\
 &-\frac{e^3}{4}\widetrp\left[\cG_{\bp,ip_0+iq_0}\cF^\alf_\bp\cG_{\bp,ip_0}\big(\partial_\gamma\sgc\big)\cF^\delta_\bp\cE^\beta_\bp\right] \label{D51+D52part2} \\
 &-\frac{e^3}{4}\widetrp\left[\cG_{\bp,ip_0+iq_0}\cF^\alf_\bp\cG_{\bp,ip_0}\sgc\big(\partial_\gamma\cF^\delta_\bp\big)\cE^\beta_\bp\right] \label{D51+D52part3}\\
 &-\frac{e^3}{4}\widetrp\left[\cG_{\bp,ip_0+iq_0}\cF^\alf_\bp\cG_{\bp,ip_0}\sgc\cF^\delta_\bp\big(\partial_\gamma\cE^\beta_\bp\big)\right] \label{D51+D52part4}\, .
\end{align}
We go through the four parts: 
(1) The derivative in \eqref{D51+D52part4} gives by definition $\partial_\gam\cE^\beta_\bp=\cE^{\beta\gam}_\bp$.
(2) The term in \eqref{D51+D52part2} containing $\partial_\gam\sgc$ cancels by the corresponding part in $(\gam\leftrightarrow\delta)$ since
$(\partial_\gam\sgc)\cF^\delta_\bp=\partial_\gam\left(\frac{1}{E^+_\bp-E^-_\bp}\right)F^\delta_\bp\sg^z\sg^x=-\frac{1}{(E^+_\bp-E^-_\bp)^2}\frac{2h_\bp}{\Delta}F^\gam_\bp F^\delta_\bp\sg^z\sg^x$,
where $\sg^x$ and $\sg^z$ are the Pauli-matrices.
(3) In order to see the cancellation of \eqref{D51+D52part3} containing $\partial_\gam\cF^\delta_\bp$ we use 
$\cF^\delta_\bp=F^\delta_\bp\sg^x=\frac{2\Delta}{E^+_\bp-E^-_\bp}h^\delta_\bp\sg^x$ defined in Eq.~\eqref{Fp}. The derivative 
of $1/(E^+_\bp-E^-_\bp)$ as well as of $h^\delta_\bp$ cancels again by the corresponding part in $(\gam\leftrightarrow\delta)$. 
(4) For \eqref{D51+D52part1} containing $\partial_\gamma \cF^\alf_\bp$ we again use the definition of $\cF^\alf_\bp$. Whereas the derivative 
of $1/(E^+_\bp-E^-_\bp)$ cancels due to the corresponding part in $(\gam\leftrightarrow\delta)$, the derivative of $h^\alf_\bp$
now produces the off-diagonal matrix of the second order vertex $\cH^{\alf\gam}_\bp= \frac{2\Delta}{E^+_\bp-E^-_\bp}h^{\alf\gam}_\bp\sg^x$
defined in Eq.~\eqref{Hp}.
Thus, the contributions finally reduce to
\begin{align} 
 \eqref{rec4part1}+\eqref{rec4part2}=&-\frac{e^3}{4}\widetrp\left[\cG_{\bp,ip_0+iq_0}\cH^{\alf\gamma}_\bp\sgc\cG_{\bp,ip_0}\cF^\delta_\bp\cE^\beta_\bp\right] \nonumber \\
  &-\frac{e^3}{4}\widetrp\left[\cG_{\bp,ip_0+iq_0}\cF^\alf_\bp\sgc\cG_{\bp,ip_0}\cF^\delta_\bp\cE^{\beta\gamma}_\bp\right] \, .
\end{align}
We commuted $\sgc$ and $\cG_{\bp,ip_0}$. We reinstall three Green functions by using the purely algebraic relation 
$\sgc\cG_{\bp,ip_0}\cF^\delta_\bp=\cG_{\bp,ip_0}\cF^\delta_\bp\cG_{\bp,ip_0}-\cF^\delta_\bp\cG_{\bp,ip_0}\sgc$.
The former part of this decomposition containing $\sgc$ cancels by the corresponding term in $(iq_0\leftrightarrow -iq_0)$ when shifting the Matsubara summation and commuting 
the matrices. We end up with identifying 
\begin{align} 
 \eqref{rec4part1}+\eqref{rec4part2}=-(\Kabgd)^{(\text{tri,IV})}+(\Kabgd)^{(\text{tri,V})}
\end{align}
in Eq.~\eqref{KH1iv} and \eqref{KH1v}.
\subsubsection{Combining $(\Kabgd)^{(\text{tri})}$ and $(\Kabgd)^{(\text{rec})}$}

We started with Eq.~\eqref{K_H}. We split $ K_{iq_0}^{\alf\beta\gam\delta} = \frac{\partial}{\partial q_\delta}
 \left. K_{\bq,iq_0}^{\alf\beta\gam} \right|_{\bq=\b0}$ into the two parts $(\Kabgd)^{(\text{tri})}$ and $(\Kabgd)^{(\text{rec})}$.
 Combining these by using the explicit matrix form, we finish with the following result:
\begin{align}
  \Kabgd = & \, (\Kabgd)^{(\text{tri})} + (\Kabgd)^{(\text{rec})} \\
  =&-\frac{e^3}{4}\trp\left[(\cG_{\bp,ip_0+iq_0}-\cG_{\bp,ip_0-iq_0})\cE^\alf_\bp\cG_{\bp,ip_0}\cE^\delta_\bp\cG_{\bp,ip_0}\cE^{\beta\gam}_\bp\right] \nonumber \\
  &-\frac{e^3}{4}\trp\left[(\cG_{\bp,ip_0+iq_0}-\cG_{\bp,ip_0-iq_0})\cF^\alf_\bp\cG_{\bp,ip_0}\cE^\delta_\bp\cG_{\bp,ip_0}\cH^{\beta\gam}_\bp\right] \nonumber \\
  &-\frac{e^3}{2}\trp\left[(\cG_{\bp,ip_0+iq_0}-\cG_{\bp,ip_0-iq_0})\cF^\alf_\bp\cG_{\bp,ip_0}\cF^\delta_\bp\cG_{\bp,ip_0}\cE^{\beta\gam}_\bp\right] \nonumber \\
  &-(\alf\leftrightarrow\beta)-(\gam\leftrightarrow\delta) \, .
\end{align}
This final result is given in Eq.~\eqref{Kabcd}.

%%%%%%%%%%%%%%%%%%%%%%%%%%%%%%%%%%%%%%%%%%%%%%%%%%%%%%%%%%%%%%%%%%%%%%%%%%%%%%%%%%%%%%%%%

\subsection{Matsubara sum and analytic continuation}

All contributions in Eq.~\eqref{Kabcd} contain a Matsubara sum of the form
\begin{equation} \label{KHiq0}
 K_{iq_0}^H = T \sum_{p_0} {\rm tr} \big[ (\cG_{ip_0+iq_0} - \cG_{ip_0-iq_0})
 \cM_1 \cG_{ip_0} \cM_2 \cG_{ip_0} \cM_3 \big] \, ,
\end{equation}
with arbitrary frequency-independent matrices $\cM_i$. Momentum dependencies are not written here.
We apply the general formula Eq.~(\ref{Imn}) in Appendix~\ref{AnalyticContinuation} for the case $m=1,n=2$ and $m=2$, $n=1$, after shifting the Matsubara summation by $q_0$. Using the relation
$\cA(\eps)\cM_2\cP(\eps)+\cP(\eps)\cM_2\cA(\eps)=-\frac{1}{2\pi i}\big(\cG^R_\eps\cM_2\cG^R_\eps-\cG^A_\eps\cM_2\cG^A_\eps\big)$,
we get
\begin{align}
 K^H_{iq_0\rightarrow\om+i0^+}=\frac{1}{2\pi i}\int d\eps \,f(\eps)\,{\rm tr}
 &\left[-\left(\cG^R_\eps-\cG^A_\eps\right)\cM_1\left(\cG^A_{\eps-\om}\cM_2\cG^A_{\eps-\om}-\cG^A_\eps\cM_2\cG^A_\eps\right)\cM_3\right.\nonumber\\
 &\left.\,+\left(\cG^R_\eps-\cG^A_\eps\right)\cM_1\left(\cG^R_{\eps+\om}\cM_2\cG^R_{\eps+\om}-\cG^R_\eps\cM_2\cG^R_\eps\right)\cM_3\right.\nonumber\\
 &\left.\,-\left(\cG^R_{\eps+\om}-\cG^R_\eps\right)\cM_1\left(\cG^R_\eps\cM_2\cG^R_\eps-\cG^A_\eps\cM_2\cG^A_\eps\right)\cM_3\right.\nonumber\\
 &\left.\,+\left(\cG^A_{\eps-\om}-\cG^A_\eps\right)\cM_1\left(\cG^R_\eps\cM_2\cG^R_\eps-\cG^A_\eps\cM_2\cG^A_\eps\right)\cM_3\right] \, .
\end{align}
For the dc Hall conductivity, we need to compute the ratio $K^H_{iq_0\rightarrow\omega+i0^+}/\om$ in the limit $\om\rightarrow 0$. We use
$(\cG^A_{\eps-\om}-\cG^A_\eps)/\omega\rightarrow -\partial_\eps\cG^A_\eps$, $(\cG^R_{\eps+\om}-\cG^R_\eps)/\omega\rightarrow \partial_\eps\cG^R_\eps$,
$(\cG^A_{\eps-\om}\cM_2\cG^A_{\eps-\om}-\cG^A_\eps\cM_2\cG^A_\eps)/\omega\rightarrow-\partial_\eps(\cG^A_\eps\cM_2\cG^A_\eps)$,
and $(\cG^R_{\eps+\om}\cM_2\cG^R_{\eps+\om}-\cG^R_\eps\cM_2\cG^R_\eps)/\omega\rightarrow\partial_\eps(\cG^R_\eps\cM_2\cG^R_\eps)$ for $\om\rightarrow 0$. Writing real and imaginary parts explicitly, one obtains Eq.~(\ref{hallcont}).

% We will use this result in the following calculations. Simple restructuring leads to the alternative form 
% %
% \begin{align*}
%  \lim_{\om \to 0} \frac{1}{\om} \bbK_{iq_0\to\om+i0^+}^H& =\frac{1}{\pi}\int d\eps \,f(\eps)\\
%  &\times \text{tr}\big[\partial_\eps\left(\Im\{\cG^R_\eps\cM_1\cG^A_\eps\cM_2\cG^A_\eps\cM_3\}\right)-\Im\{(\partial_\eps\cG^R_\eps)\cM_1\cG^R_\eps\cM_2\cG^R_\eps\cM_3-\cG^R_\eps\cM_1\partial_\eps(\cG^R_\eps\cM_2\cG^R_\eps)\cM_3\}\big]
% \end{align*}
%  \end{widetext}
 
\end{appendix}
\twocolumngrid
%%%%%%%%%%%%%%%%%%%%%%%%%%%%%%%%%%%%%%%%%%%%%%%%%%%%%%%%%%%%%%%%%%%%%%%%%%


\begin{thebibliography}{apsrev}
\bibitem{footnote1} The quasi-particle spectral function must not be
confused with the spectral function of the electrons appearing in the
spectral representation of the Green function defined with bare
(unrotated) electron operators.
\bibitem{footnote2} For symmetric matrices $M_1$ to $M_m$ one can easily prove the following equality under the trace: 
$\text{tr}\big(M_1\dots M_m\big)=\text{tr}\big(M_m\dots M_1\big)$. All $2 \times 2$ matrices we consider (e.g., $\cE^\alf_\bp$, $\cF^\alf_\bp,\dots$)
in this paper are symmetric. In the text, we refer to this kind of manipulation as reversing the matrix order under the trace.
\end{thebibliography}
\end{document}